\documentclass[10pt,secnumarabic,amssymb, nobibnotes, aps]{revtex4-2}
\usepackage{amsmath}
\usepackage{amssymb}
\usepackage{amsthm}
\usepackage{bbm}
\usepackage{graphicx}
\usepackage{hyperref}
\usepackage{physics}
\usepackage{systeme}
\usepackage{times}
\usepackage{epstopdf}
\usepackage{float}
\usepackage{xcolor}
\usepackage[normalem]{ulem}

\hypersetup{
	colorlinks=true,linkcolor=blue,citecolor=blue,
	filecolor=blue,urlcolor=blue,breaklinks=true
}

\begin{document}
\title{Approximate analytical solutions of the Schr\"{o}dinger equation with Hulth\'{e}n potential in the global monopole spacetime}

\author{Saulo S. Alves}
\email{marciomc05@gmail.com}
\affiliation{
        Departamento de F\'{i}sica,
        Universidade Federal do Maranh\~{a}o,
        65085-580, S\~{a}o Lu\'{i}s, Maranh\~{a}o, Brazil
      }
\author{M\'{a}rcio M. Cunha}
\email{marcio.cunha@penedo.ufal.br; marciomc05@gmail.com }
\affiliation{
        Unidade Educacional de Penedo, Campus Arapiraca, Universidade Federal de Alagoas, Av. Beira Rio, s/n - Centro Histórico, 57200-000 Penedo, Alagoas, Brazil
      }
  \author{Hassan Hassanabadi}
  \email{h.hasanabadi@shahroodut.ac.ir}
  \affiliation{
  	Faculty of Physics, Shahrood University of Technology, Shahrood, Iran\\
  	$^{4}$ \quad Department of Physics, University of Hradec Kr\'{a}lov\'{e},
  	Rokitansk\'{e}ho 62, 500 03 Hradec Kr\'{a}lov\'{e}, Czech Republic
  }
\author{Edilberto O. Silva}
\email{edilberto.silva@ufma.br; edilbertoo@gmail.com}
\affiliation{
        Departamento de F\'{i}sica,
        Universidade Federal do Maranh\~{a}o,
        65085-580, S\~{a}o Lu\'{i}s, Maranh\~{a}o, Brazil
      }
\date{\today }
\begin{abstract}
In this paper, we study the nonrelativistic quantum mechanics of an electron in a spacetime containing a topological defect. We also consider that the electron is influenced by the Hulth\'{e}n potential. In particular, we deal with the Schr\"{o}dinger equation in the presence of a global monopole. We obtain approximate solutions for the problem, determine the scattering phase shift and the $S$-matrix, and analyze bound states.
\end{abstract}

\maketitle

\section{Introduction}
\label{intro}

The search for exactly solvable models is one of the most relevant tasks in the branch of theoretical physics.
However, obtaining exact solutions is not possible in all cases of research interest. In that circumstances, it is necessary to study how to make corrections to these models or even to examine the obtaining of approximate solutions.
In the framework of nonrelativistic quantum mechanics, the Schr\"{o}dinger equation describes precisely the dynamics of a closed quantum system, providing all the information about the properties of the system \cite{PP.2012.182}. In this context, several examples of exactly solvable models, like the harmonic oscillator \cite{Book.2005.Griffiths}, the two-body problem with non-central forces \cite{AJP.1963.16.311}, the modified ring-shaped oscillator potential \cite{EPJP.2022.137.527}, and a model involving a class of hyperbolic potential well \cite{CPB.2022.31.040301}.
Because of its fundamental feature, it is possible to think about the Schr\"{o}dinger equation in the most diverse contexts, describing low-dimensional electron gases \cite{PhysRevB.48.1626}, problems with anisotropic mass \cite{10.1016/j.physe.2021.114827}, as well as the presence of curvature and torsion in the spacetime \cite{10.1209/0295-5075/132/10004}. Concerning the Schr\"{o}dinger equation in a curved space, a relevant type of problem consists of studying the presence of topological defects \cite{Book.2000.Vilenkin}.
Topological defects can arise in the contexts of gravitation and condensed matter physics.  In the first one, topological defects are associated with the process of evolution in the early universe, in which symmetry-breaking phase transitions took place \cite{10.1016/S0370-1573(02)00014-5}.  In condensed matter physics, these defects might appear during material synthesis, being inevitable in the process \cite{doi:10.1021/acs.jpclett.8b03225}.
The study of the physical implications of these defects on the physical properties of a system has been an active research line in the last decades.

In gravitation, we can cite topological defects such as cosmic strings, domain walls, and global monopoles \cite{Book.2000.Vilenkin,book.Jagodzinski.1984,PRD.1981.23.852}.
As pointed out in \cite{10.1088/0264-9381/19/5/310}, cosmic strings and global monopoles are exotic topological defects. Around a cosmic string, spacetime is not globally flat but locally flat. Another remarkable feature about cosmic strings refers to the fact we can associate a conical geometry to this type of defect. Because of this, there is a counterpart for cosmic strings in condensed matter physics, which are named disclinations \cite{Book.katanaev.TD.2005,Katanaev19921}. A disclination also presents a conical geometry that manifests in solids and liquid crystals. The spacetime of cosmic string exerts an impressive influence on the behavior of a physical system, in which effects such as gravitational leasing, self-force, and the gravitational Aharonov-Bohm effect take place. In the literature,
different models have been solved in the cosmic string spacetime \cite{IJMPA.2022.37.2250122,EPJC.2022.82.688,RP.2022.39.105749,IJGMMP.2022.19.2250133,CPC.2022.46.043104,AOP.2022.440.168857,CQG.2022.39.075006,CQG.2022.39.075007,CJP.2021.99.496,IJGMMP.2021.18.2150187,CQG.2021.38.205006,FBS.2021.62.57,IJMPE.2021.30.2150050,MPLA.2021.36.2150004,universe.2020.6.6110203,PRD.2020.102.105020,IJMPA.2015.30.1550124} and disclinations \cite{PRL.2021.127.066401,AoP.2021.425.168384,IJMPA.2020.35.2050071,EPJB.2021.94.75,PNAS.2022.119.e2122226119,RSC.2022.24.15691,IJMPA.2022.37.2250046,PRE.2021.104.065002}, which demonstrates the wide range of possibilities of investigation involving these issues.

A global monopole also presents significant features, like the topological scattering of test particles. For this defect, one can say its influence manifests because of a solid deficit angle that dictates the corresponding spacetime's topological behavior and curvature.
In condensed matter physics, topological defects might appear in various scenarios. For instance, as domain walls in magnetic materials \cite{Book.Kleinert.GFCM.1989}, vortices in superconductors \cite{doi:10.1142/0356} and solitons in polyacetylene \cite{PhysRevB.22.2099}. Besides, there are defects known as dislocations, which occur in disordered solids \cite{RMP.2008.80.61,ACSA.1984.40.309}.  In the present manuscript, we are interested in studying the influence of a spacetime produced by a global monopole on the quantum mechanics of a nonrelativistic particle. For a global monopole, the scalar matter field plays the role of an order parameter that, outside the monopole, the core acquires a non-vanishing value. In this context, Barriola and Vilenkin \cite{PRL.1989.63.341} presented an approximate solution of the Einstein equations \cite{doi:10.1080/17455030.2019.1574410, 10.1088/1674-1056/22/10/100203,10.1088/0031-8949/87/03/035003} for the metric outside a monopole resulting from the global $O(3)$ symmetry breaking. They showed that the monopole exerts practically no gravitational
force on the nonrelativistic matter. Besides, the space around it has a solid angle deficit, and the same angle, independent of the impact parameter, deflects all light rays. After this study, several works have been developed in this direction, which include the study of the polarization and vacuum fluctuations  \cite{PRD.2022.105.085006,JETP.2016.123.807,PRD.2018.98.065009}, the relativistic motion of quantum oscillator \cite{SR.2022.12.8794,EPJP.2021.137.54,EPJC.2020.80.206}, radiation, absorption and scattering of black holes \cite{PLB.2019.788.231,IJTP.2017.56.2061,CPB.2012.21.040404}, etc.

As we have mentioned, the study of quantum mechanics problems in which the Schr\"{o}dinger equation has no exact solutions is a question of interest due to its foundational character. To deal with these problems, it is often necessary to implement approximation techniques. As examples of investigation in this context, we can cite solutions of the Schr\"{o}dinger equation with Eckart plus inversely quadratic Yukawa potentials \cite{JMM.2020.26.349}, Hua plus modified Eckart potential with the centrifugal term \cite{EPJP.2020.135.571}, Manning-Rosen plus Hellmann potential and its thermodynamic properties \cite{EPJP.2019.134.315}, and shifted Deng–Fan potential model \cite{MP.2014.112.127}.
Then, studying the nonrelativistic quantum motion of a particle in the presence of both the Hulth\'{e}n potential and the global monopole is a pertinent issue.

Having all this in mind, in this manuscript, our goal consists of obtaining approximate solutions for the Schrodinger equation in the spacetime of a global monopole in the presence of the Hulth\'{e}n potential. That potential class presents an extensive scope of applications, such as nuclear physics, chemical physics, and condensed matter systems \cite{10.1088/0031-8949/76/1/016}. Regarding the study of approximate solutions when the Hulth\'{e}n potential is present, we can cite examples of works in the literature dealing with this topic in the framework of the Schr\"{o}dinger equation \cite{10.1088/0031-8949/80/06/065304}, the Klein-Gordon equation \cite{10.1088/0031-8949/76/6/005}, and Dirac equation \cite{10.1016/j.physleta.2005.12.039,EPJC.2017.77.270}.

The organization of the paper is as follows. In Section \ref{sec2}, we write down the Schr\"{o}dinger equation for a charged particle in the global monopole spacetime, including the Hulth\'{e}n potential. Before solving the radial equation, we study the effective potential to which the particle is subjected. We show that bound states are allowed for $\alpha<1$ and $\alpha>1$, and the other parameters are fixed. To solve the radial equation of motion for arbitrary states, we use standard approximations to the Hulth\'{e}n potential. We show that the radial differential equation is of the hypergeometric type, and its solution is given in terms of the hypergeometric 
function $_{2}{F}_{1}\left(a,b;\,c;\,z\right)$. In Section \ref{sec3}, we solve the radial equation for scattering states and find the phase shift of the wave function. Section \ref{sec4} is dedicated to examining bound states solutions. To do this, we first determine the $S$-matrix and then analyze its poles, from which an expression for the energy of bound states can be obtained. We end this section by presenting some sketches for the energy levels and comparisons with other results in the literature. Finally, we make our concluding remarks on Section \ref{sec5}.

\section{Nonrelativistic quantum mechanics in the global monopole spacetime}\label{sec2}

\begin{figure}[h]
	\centering
	\includegraphics[scale=0.33
	]{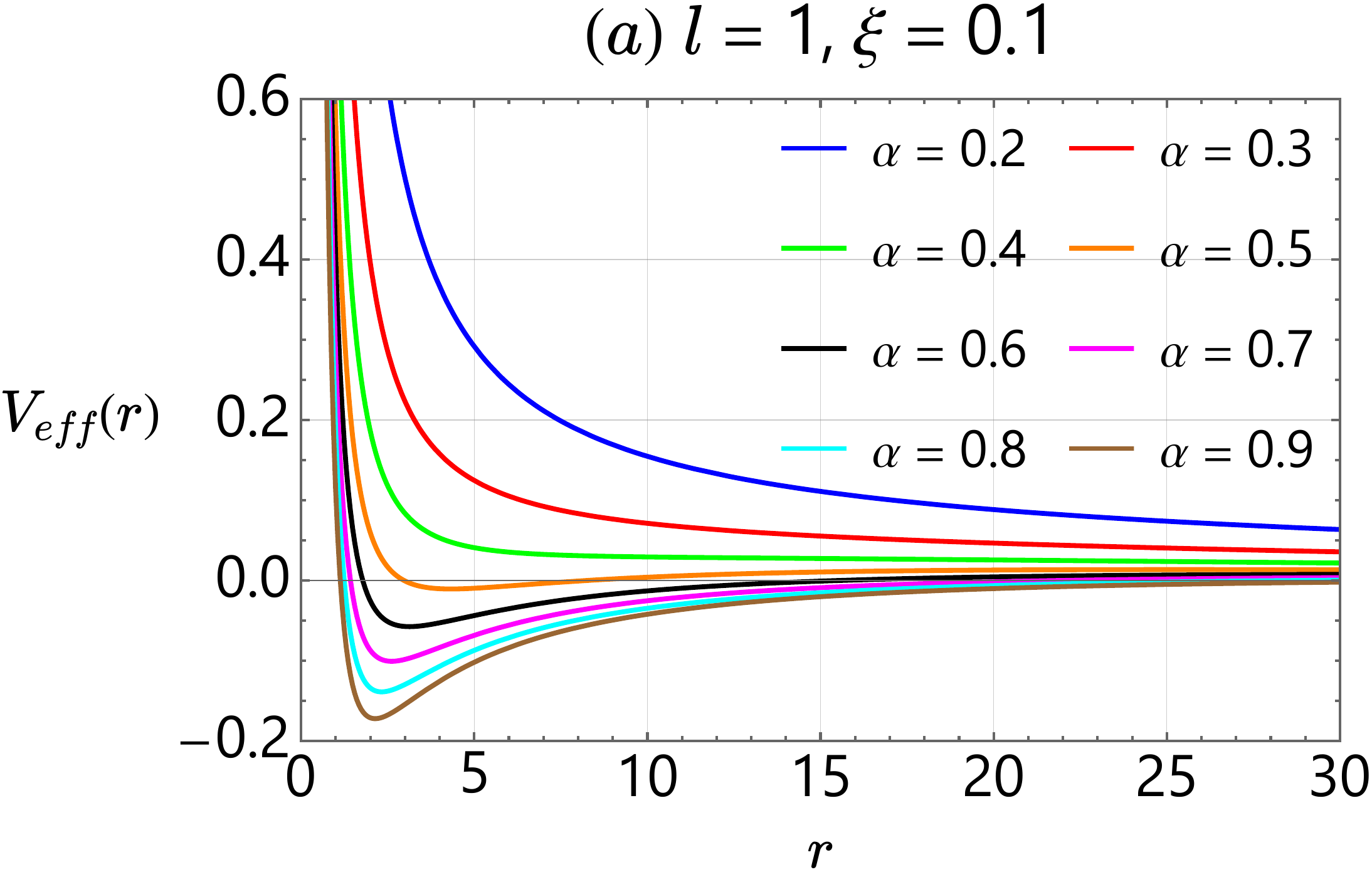}\qquad
	\includegraphics[scale=0.33
	]{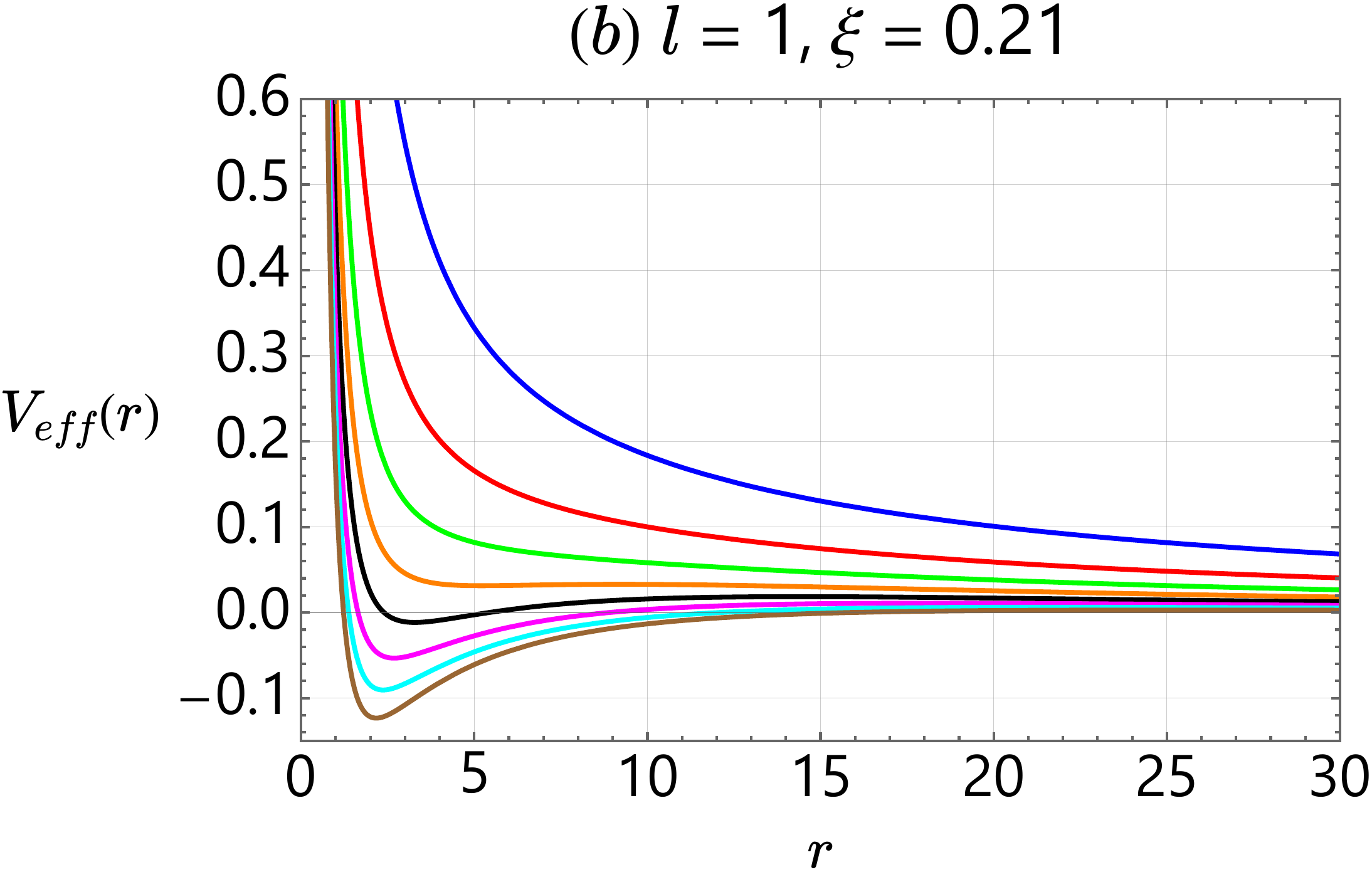}
	\includegraphics[scale=0.33
	]{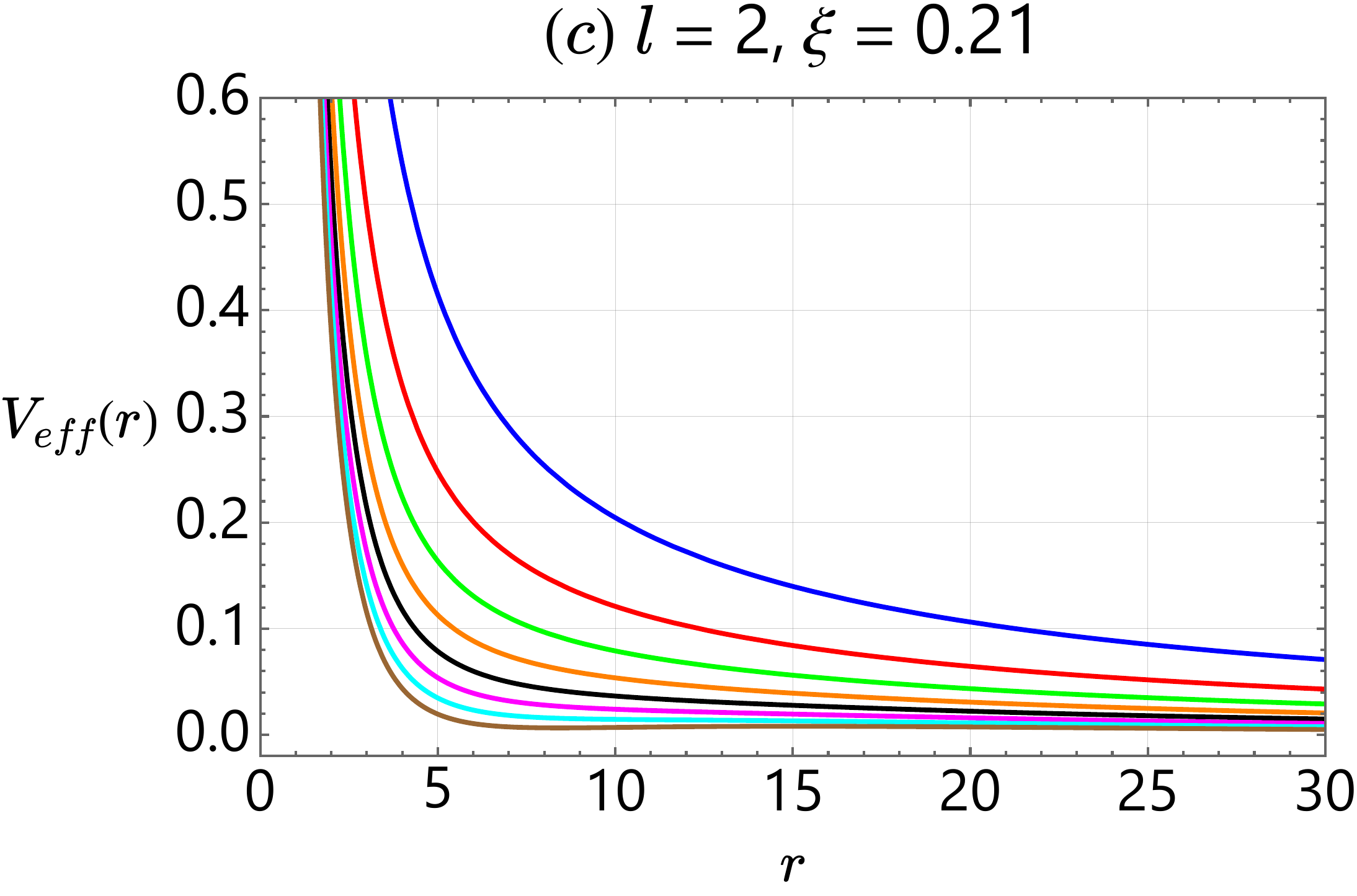}
	\includegraphics[scale=0.33
	]{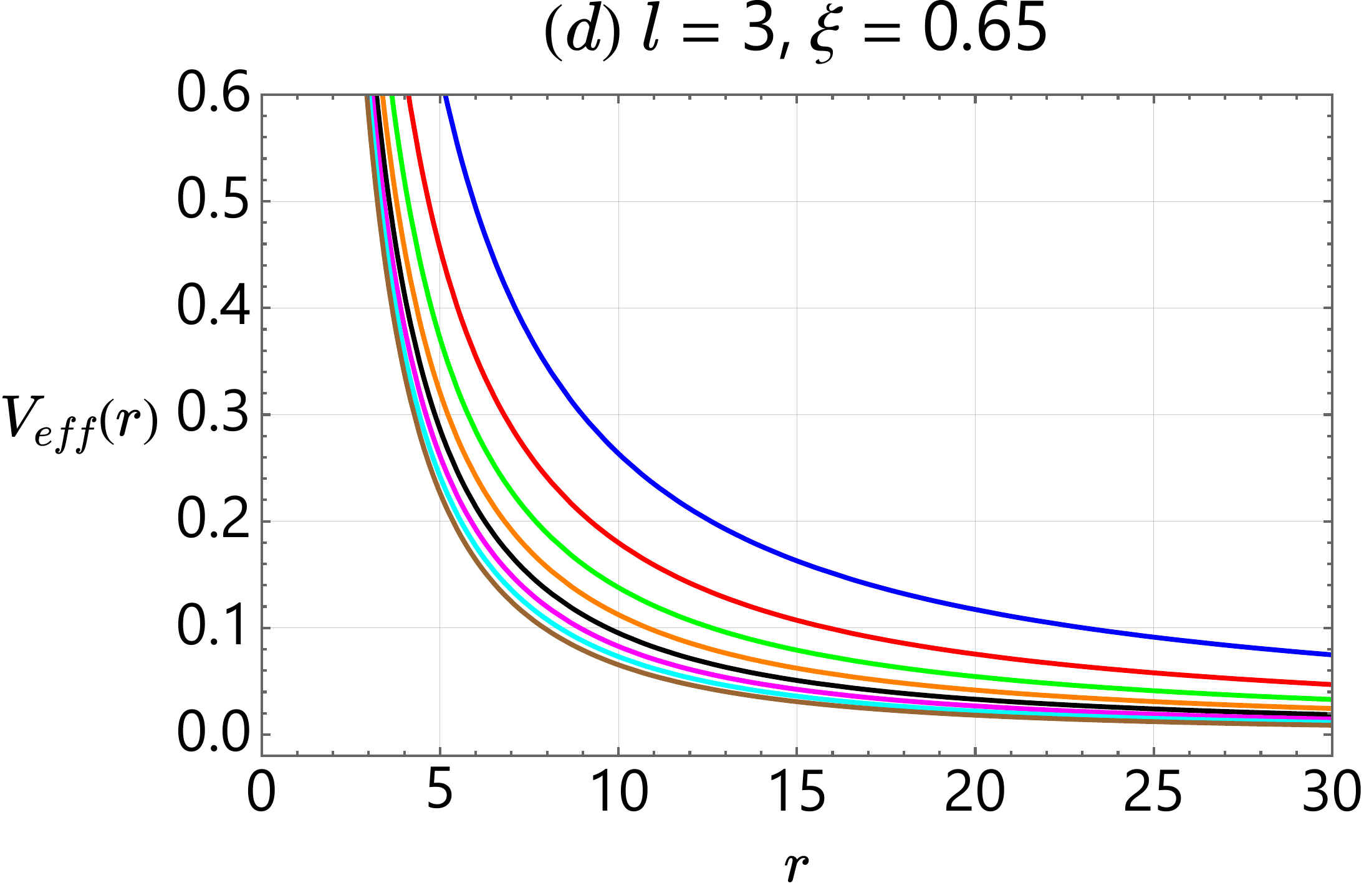}
	\caption{Effective potential (Eq. (\ref{veff})) as a function of $r$ for different values of $\alpha<1$. Four situations involving $\xi$ and $l$ are considered: (a) $\xi=0.1$ and $l=1$, (b) $\xi=0.65$ and $l=1$, (c) $\xi=0.21$ and $l=2$, and (d) $\xi=0.65$ and $l=3$.}
	\label{Fig_veff}
\end{figure}

\begin{figure}[!t]
	\centering
	\includegraphics[scale=0.33
	]{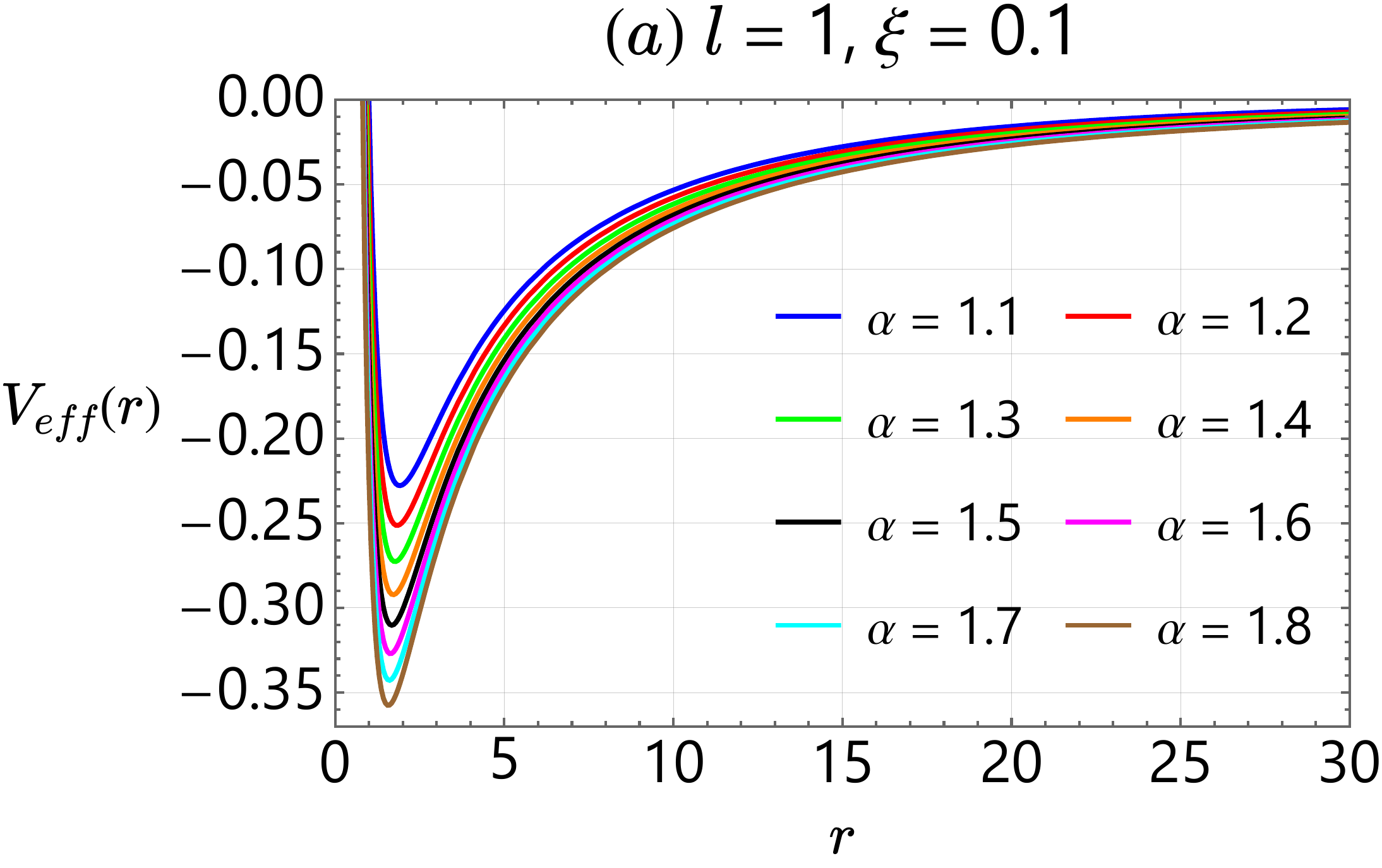}\qquad
	\includegraphics[scale=0.33
	]{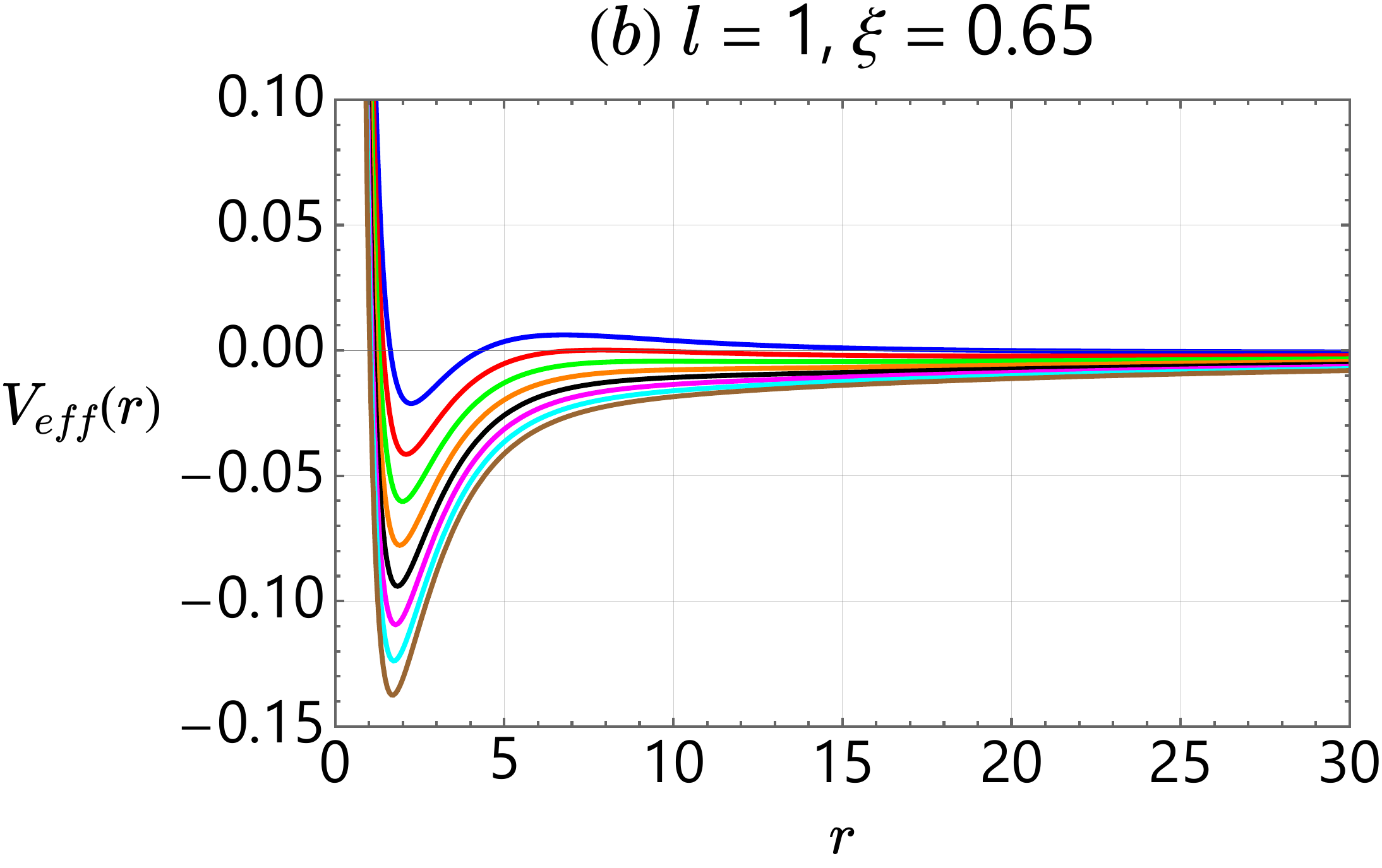}
	\includegraphics[scale=0.33
	]{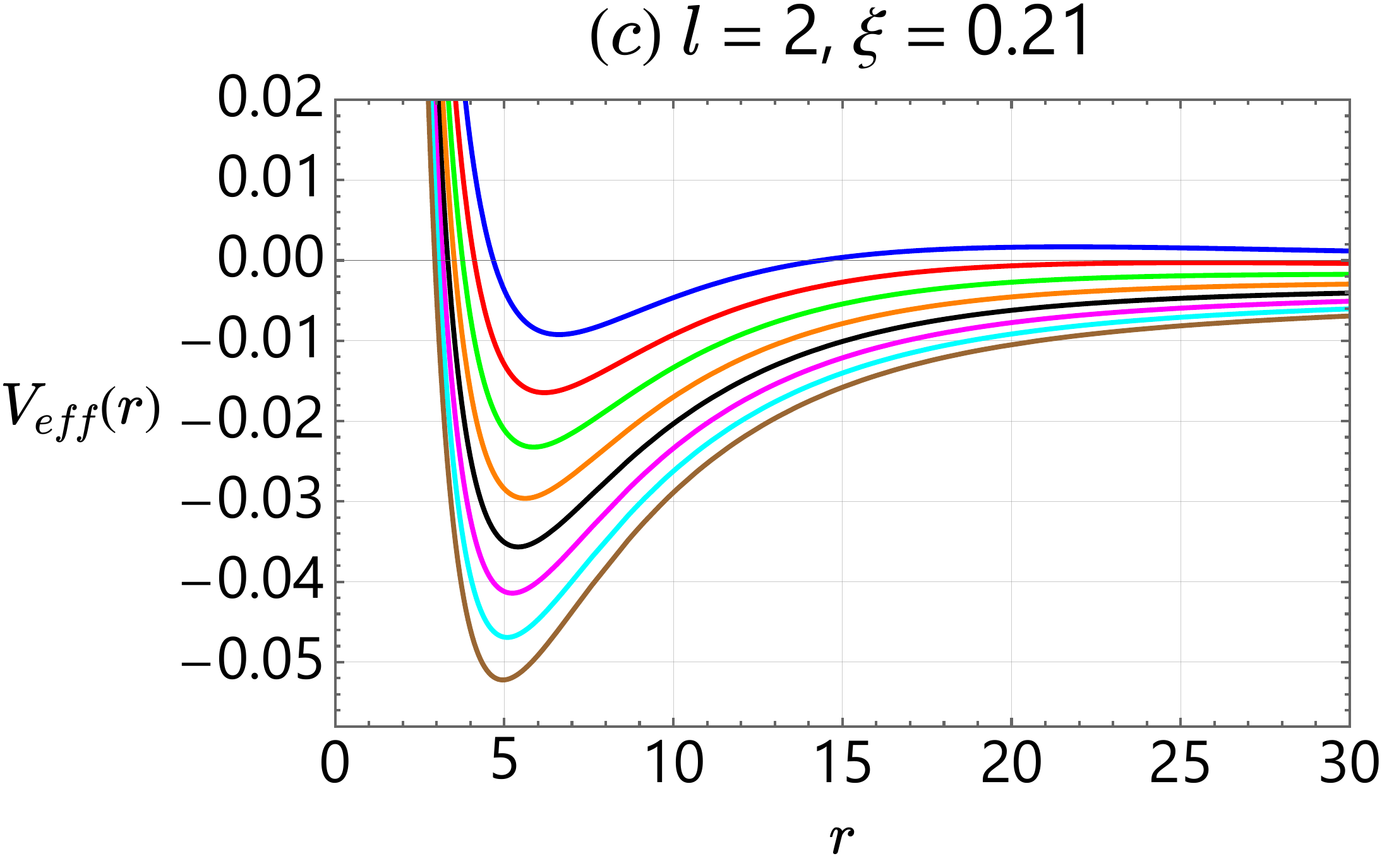}
	\includegraphics[scale=0.33
	]{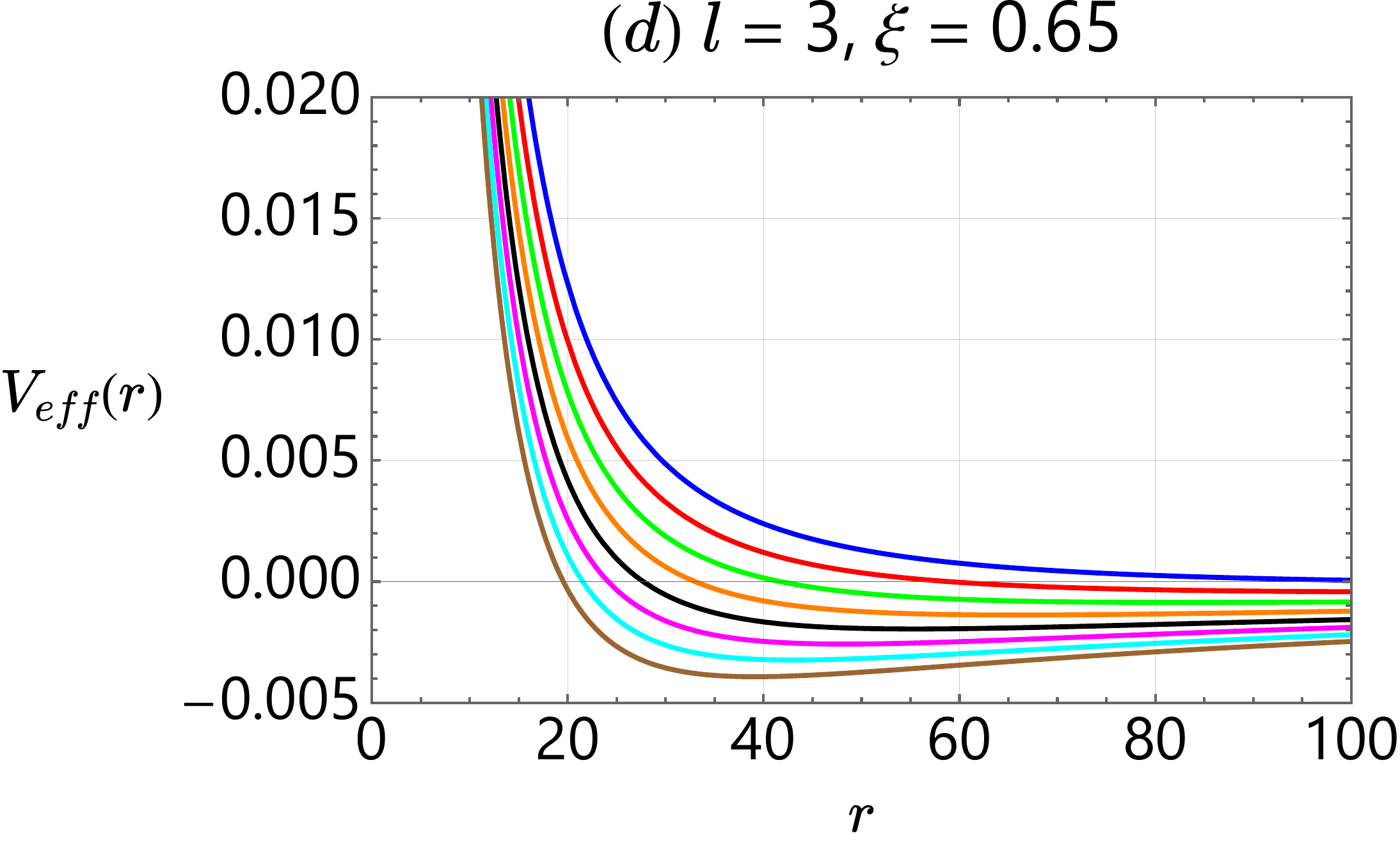}
	\caption{Effective potential (Eq. (\ref{veff})) as a function of $r$ for different values of $\alpha>1$. Four situations involving $\xi$ and $l$ are considered: (a) $\xi=0.1$ and $l=1$, (b) $\xi=0.65$ and $l=1$, (c) $\xi=0.21$ and $l=2$, and (d) $\xi=0.65$ and $l=3$.}
	\label{Fig_veff2}
\end{figure}

In this section, we write down the Schr\"{o}dinger equation with vector coupling
to describe the motion of an electron interacting with the Hulth\'{e}n
potential in the global monopole spacetime. The metric of this manifold is
expressed by the line element \cite{PRL.1989.63.341}
\begin{equation}
	ds^{2}=-dt^{2}+\alpha ^{-2}dr^{2}+r^{2}\left( d\theta ^{2}+\sin ^{2}\theta
	d\varphi ^{2}\right) ,  \label{metric}
\end{equation}
where $\alpha ^{2}=1-8\pi G\eta ^{2}$ is smaller than $1$, and represents the deficit solid of this
manifold. The parameter $\eta$ corresponds to the scale of gauge-symmetry breaking \cite{PhysRevD.95.104012}. For this spacetime, it is known that the area of a sphere of unit radius is not $4\pi r^{2}$, but $4\pi \alpha^{2}r^{2}$. Furthermore, the surface with $\theta=\pi/2$ presents the geometry of a cone (gauge cosmic string) with a deficit angle $\Tilde{\Delta}= 8\pi^{2}\eta^{2}$. 
It is known that the motion of massive or charged particles in the spacetime (\ref{metric}) involves the effect of the self-interaction potential in the model description. In this case, the relevant equation is the Schr\"{o}dinger equation in spherical polar coordinates with vector
coupling. The Schr\"{o}dinger equation has the form
\begin{equation}
	H\psi =E\psi ,  \label{SE}
\end{equation}
where $H$ is the corresponding Hamiltonian operator, given by
\begin{equation}
	H=\frac{1}{2M}\mathbf{p}^{2} +V_{eff}\left( r\right).  \label{SH}
\end{equation}
Here, $\mathbf{p}$ denotes the momentum operator, $M$ is the mass of the particle and
\begin{equation}
	V_{eff}\left( r\right) =V_{H}\left( r\right) +V_{SI}\left( r\right)  \label{vr}
\end{equation}
is the effective potential, which contains the Hulthén potential $V_{H}\left( r\right) $ and the electrostatic self-interaction potential $V_{SI}\left( r\right)$.
The self-interaction potential $V_{SI}\left( r\right)$ is of the  Coulomb-type, given by \cite {PRD.1997.56.1345}
\begin{equation}
	V_{SI}(r)=\frac{\mathcal{K}\left( \alpha \right) }{r},  \label{si}
\end{equation}
where $r$ is the distance from the electron to the monopole and $\mathcal{K}(\alpha)$
is the constant of coupling. In a general analysis, depending on the sign of $\mathcal{K}(\alpha)$, the potential $V_{SI}(r)$ can be either attractive or repulsive. For the electrostatic case, $\mathcal{K}(\alpha)$ is known to be
\cite{PRD.1997.56.1345}
\begin{equation}
	\mathcal{K}\left( \alpha \right) =\frac{e^{2}S(\alpha )}{2}>0, \label{ic}\\
\end{equation}
where $e$ is the electron charge. The function $S(\alpha)$ is given by
\begin{equation}
	S(\alpha )=\sum\limits_{l=0}^{\infty }\left[ \frac{2l+1}{\sqrt{4l\left(
			l+1\right) +\alpha ^{2}}}-1\right],  \label{sb}
\end{equation}
where $l$ denotes the angular-momentum quantum number. Note that $S(\alpha)$ is a finite positive number for $\alpha <1$ and negative for $\alpha >1$. In our approach, we consider these two possibilities for $\alpha$, and discuss the physical implications when $V_{SI}(r)$ is added to the Hulthén potential in its approximate form that we present later. Although the case with $\alpha>1$ is nonphysical, we consider it important because it makes our study more complete from a mathematical physics point of view. The Hulth\'{e}n potential is given by  \cite{AMAFA.1942.28.5,PLA.2007.368.13} 
\begin{equation}
	V_{H}\left( r\right) =-\frac{Ze^{2}\xi e^{-\xi r}}{1-e^{-\xi r}},  \label{Hp}
\end{equation}
where $Z$ is a positive constant, and $\xi$ is the screening parameter, determining the Hulthén potential range. When the potential $V_{H}(r)$
is used for atoms phenomena, the constant $Z$ is identified
with the atomic number. Some authors have opted to define a new parameter, namely $V_{0}=Ze^{2}\xi$, and state that $V_{0}$ is related to the atomic number $Z$ and the screening parameter $\xi$ \cite{CEJP.2008.6.884}. Here, we prefer explicitly conducting these parameters in our calculations as in Reference \cite{PLA.372.2008.4779}. 
With the inclusion of the effective potential (\ref{vr}), Equation (\ref{SE}) can be written explicitly as
\begin{equation}
	-\frac{\hbar ^{2}}{2M}\left[ \frac{\alpha ^{2}}{r^{2}}\frac{\partial }{\partial r}\left( r^{2}\frac{\partial }{\partial r}\right) -\frac{\mathbf{L}^{2}}{r^{2}}\right] \psi \left( \mathbf{r}\right) +\frac{\mathcal{K}\left(\alpha \right) }{r}\psi \left( \mathbf{r}\right) -\frac{Ze^{2}\xi e^{-\xi r}
	}{1-e^{-\xi r}}\psi =E\psi \left( \mathbf{r}\right),  \label{sep}
\end{equation}
where $\mathbf{L}$ is the usual orbital angular momentum operator in spherical polar coordinates. At this point, it is useful to notice that 
$\left[ H,\mathbf{L}\right] =\left[ H,\mathbf{L}^{2}\right] =0$. From these commutation relations,  together with Equation (\ref{SE}), we can write the following equations \cite{Book.1994.Sakurai} involving the angular momentum operator and its corresponding quantum numbers:
\begin{align}
	\mathbf{L}^{2}\psi \left( \mathbf{r}\right)  &=l\left( l+1\right) \hbar^{2}\psi \left( \mathbf{r}\right), \\
	L_{z}\psi \left( \mathbf{r}\right)  &=m\hbar \psi \left( \mathbf{r}\right) .
\end{align}
In this way, we shall search for energy eigenstates of the form 
\begin{equation}
	\psi \left( \mathbf{r}\right) =R_{n}\left( r\right) Y_{l}^{m}\left( \theta,\varphi \right),  \label{sols}
\end{equation}
in which $Y_{l}^{m}\left( \theta,\varphi \right)$ represents the spherical harmonics functions.
Substituting this solution into Equation (\ref{sep}) together with the standard
change of variables $R\left( r\right) =r^{-1}u(r)$, we obtain the radial equation
\begin{equation}
	-\frac{\hbar^{2} \alpha^{2}}{2M}\frac{d^{2}u(r)}{dr^{2}}+V_{eff}u(r) =Eu(r).  \label{et}
\end{equation}
where
\begin{equation}
	V_{eff}(r)= \frac{\hbar ^{2}}{2M}\frac{l\left( l+1\right) }{r^{2}} +\frac{\mathcal{K}\left( \alpha \right) }{r} -\frac{Ze^{2}\xi e^{-\xi r}}{\left( 1-e^{-\xi r}\right)}. \label{veff}
\end{equation}
and $\mathcal{K}\left( \alpha \right) $ is given in Equation (\ref{ic}). Equation (\ref{et}) is the Schr\"{o}dinger equation in the spacetime of a global monopole in the presence of the Hulthén potential. At this point, if we let off the Hulthén potential, we obtain the  Schr\"{o}dinger equation in the global monopole background.
The effective potential (\ref{veff}) allows us to study both bound and scattering states. This can be accomplished by controlling the parameters $\mathcal{K}(\alpha)$ and $\xi$, where we can observe regions of minimum potential energy for different values of $\alpha <1$ (Figure \ref{Fig_veff}). In particular, for $\xi=0.1$ and $l=1$ the system admits bound states for all $\alpha$ values considered (Figure \ref{Fig_veff}(a)). For fixed $l$, the number of curves that admit bound states decreases as the parameter $\xi$ increases (Figure \ref{Fig_veff}(b)). On the other hand, by increasing $l$, the system admits only scattering states (Figures \ref{Fig_veff}(c)-(d)). This result is due to the cancellation between the self-interaction potentials and the Hulthén potential with its approximations adopted here. A curious feature is manifested when we adopt the same parameters used in Figure \ref{Fig_veff}, but now taking $\alpha>1$. All potential minima appear in the effective potential's negative range, and bound states are permissible (Figure \ref{Fig_veff2}). The potential well gets deeper for fixed $l$ and decreasing values of $\xi$ (Figure \ref{Fig_veff2}(a)-(b)).

To solve the Equation (\ref{et}) for scattering states ($E>0$), it is convenient to analyze its asymptotic behavior for both regimes of large and small $r$. Through this
analysis, we obtain the limiting forms \cite{CPB.2009.18.3663,CEJP.2008.6.884,APPA.2013.07.20}
\begin{eqnarray}
	u(r) &\rightarrow &r^{\frac{1}{2}+\ell },\;\;r\rightarrow 0;  \label{bc} \\
	u(r) &\rightarrow &\sin \left( k r-\frac{
		\ell \pi }{2}+\delta _{\theta }\right) ,\;\;r\rightarrow \infty, \label{bcc}
\end{eqnarray}
where $\ell =\sqrt{4l^{2}+4l+\alpha ^{2}}/2\alpha$, $k^{2}=2ME/\alpha ^{2}\hbar ^{2}$. Note that if $\alpha=1$, we have $\ell  \rightarrow l+1/2$, and the limiting forms (\ref{bc}) and (\ref{bcc}) recover the usual ones \cite{landau1981quantum}. Equation (\ref{et}) cannot be
analytically solved for $l\neq 0$, even for the $s$-wave case.
Then, we must make Equation (\ref{et}) solvable for any $l$. At the same time, we want to consider the self-interaction potential (\ref{si}), which is a Coulomb-type potential. Thus, we follow the literature and proceed by taking the following approximations \cite{PRA.1976.14.2363,RP.2019.14.102409,PLA.2019.383.3010,JLTP.2021.203.84,PLA.372.2008.4779}: 
\begin{equation}
	\frac{1}{r^{2}}\approx \frac{\xi ^{2}e^{-\xi r}}{\left( 1-e^{-\xi r}\right)
		^{2}},\;\;\frac{1}{r}\approx \frac{\xi e^{-\xi r}}{1-e^{-\xi r}},
\end{equation}
which are valid only for small values of the parameter $\xi$. By defining the new variable $y=1-e^{-\xi r}$, these approximations become 
\begin{equation}
	\frac{1}{r^{2}}\approx \frac{\xi ^{2}\left( 1-y\right) }{y^{2}},\;\;\frac{1}{r
	}\approx \frac{\xi \left( 1-y\right) }{y}.
\end{equation}
Substituting the approximations above in Equation (\ref{et}) and performing the appropriate algebraic manipulations, we obtain the differential equation
\begin{equation}
	\left( 1-y\right) ^{2}\frac{d^{2}u\left( y\right) }{dy^{2}}-\left(
	1-y\right) \frac{du\left( y\right) }{dy}-\frac{\lambda ^{2}\left( 1-y\right) 
	}{y^{2}}u\left( y\right) +\frac{\wp ^{2}\left( 1-y\right) }{y}u\left(
	y\right) +\kappa^{2}u\left( y\right) =0, \label{edoH}
\end{equation}
with 
\begin{equation}
	\lambda ^{2}=\frac{l\left( l+1\right) }{\alpha ^{2}}\geq 0,\;\;\wp ^{2}=\frac{2M}{
		\hbar ^{2}\alpha ^{2}\xi }\left( Ze^{2}-\mathcal{K}\left(\alpha \right)\right)
	,\;\;\kappa =\frac{k}{\xi}.
\end{equation}
It can be shown that Equation (\ref{edoH}) has regular singularities at points $0$, $1$, and $\infty$. By a suitable change of variables, it can be converted to a hypergeometric differential equation whose solution is given in terms of the hypergeometric 
function $_{2}{F}_{1}\left( a,b;\,c;\,y\right)$. In this way, we take the wave function in the range $0\leq y\leq1$ of the form
\begin{equation}
	u\left( y\right) ={y}^{d}\left( 1-y\right) ^{-i\kappa }\,_{2}{F}_{1}\left( a,b;\,c;\,y\right), \label{gs}
\end{equation}
where the parameters $a,b,c$ and $d$ are given by
\begin{align}
	a& ={d-i\kappa +\Delta },\\
	b& ={d-i\kappa -\Delta },\;\;{\Delta =\sqrt{\wp ^{2}-\kappa ^{2}}},\label{bb}\\
	c& =2d,\;\;
	d=\frac{1}{2}\left( 1+\sqrt{1+4\,{\lambda }^{2}}\right). \label{rd}
\end{align}
\begin{figure}[!h]
	\centering
	\includegraphics[scale=0.25
	]{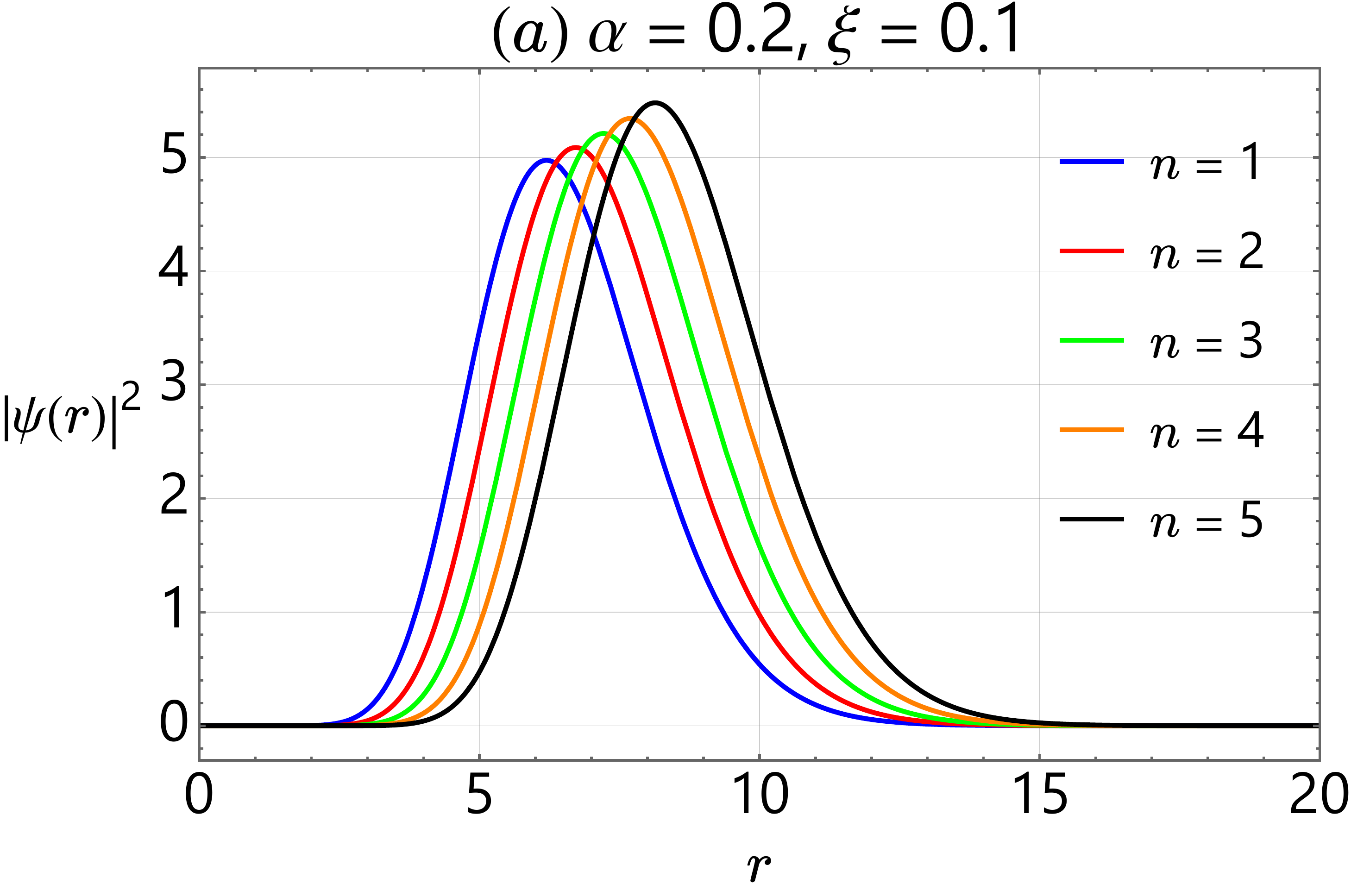}\qquad
	\includegraphics[scale=0.25
	]{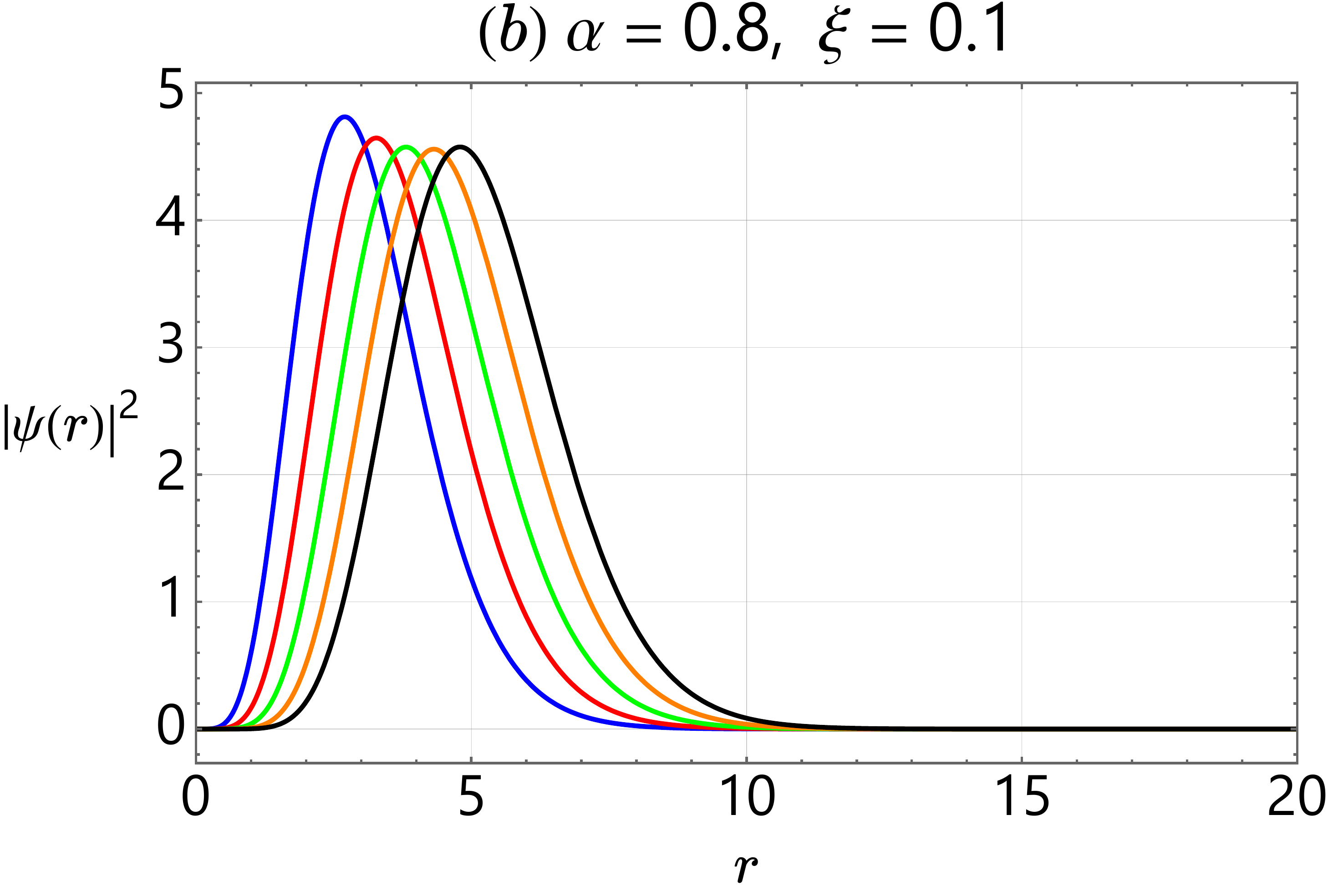}
	\includegraphics[scale=0.25
	]{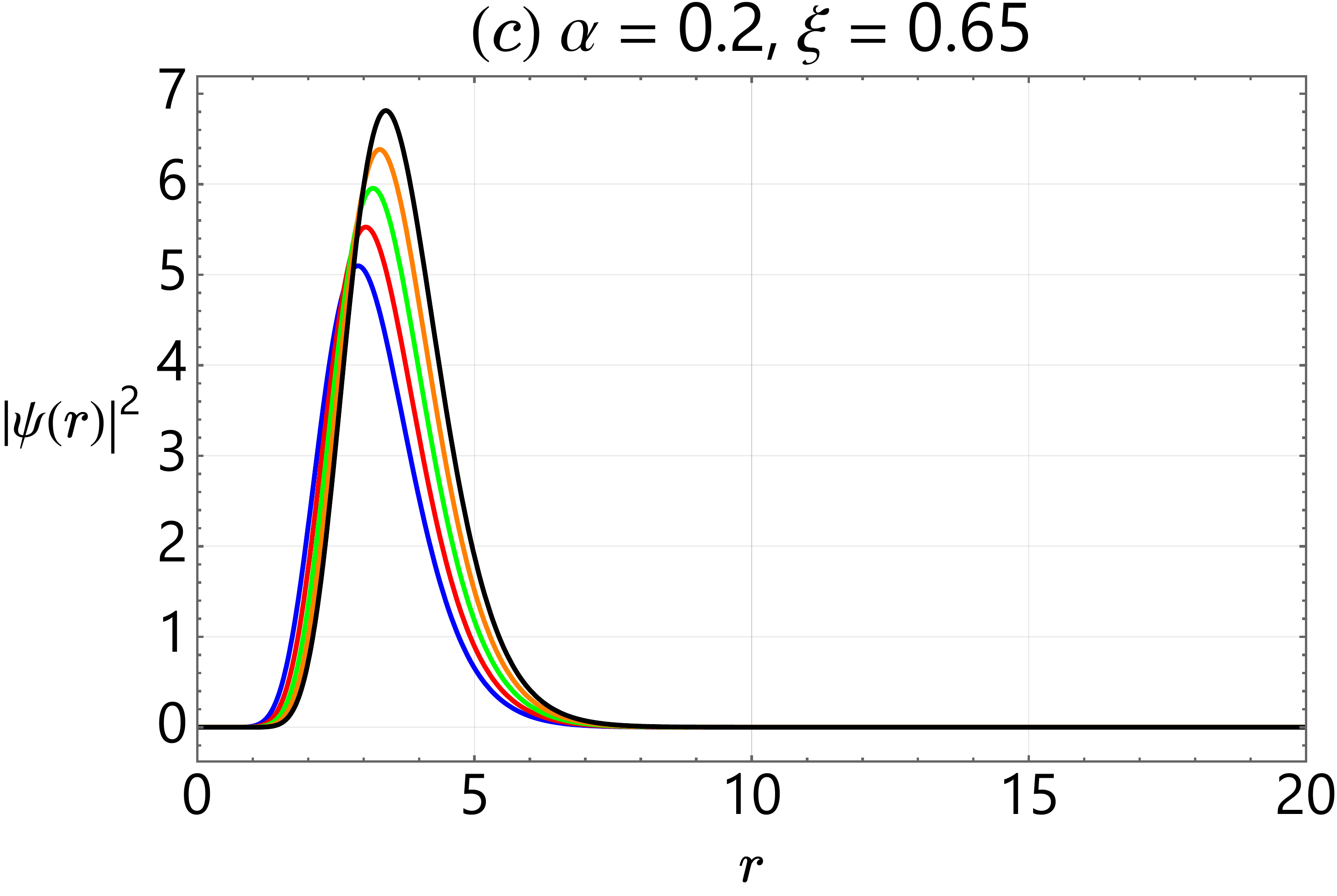}\qquad
	\includegraphics[scale=0.25
	]{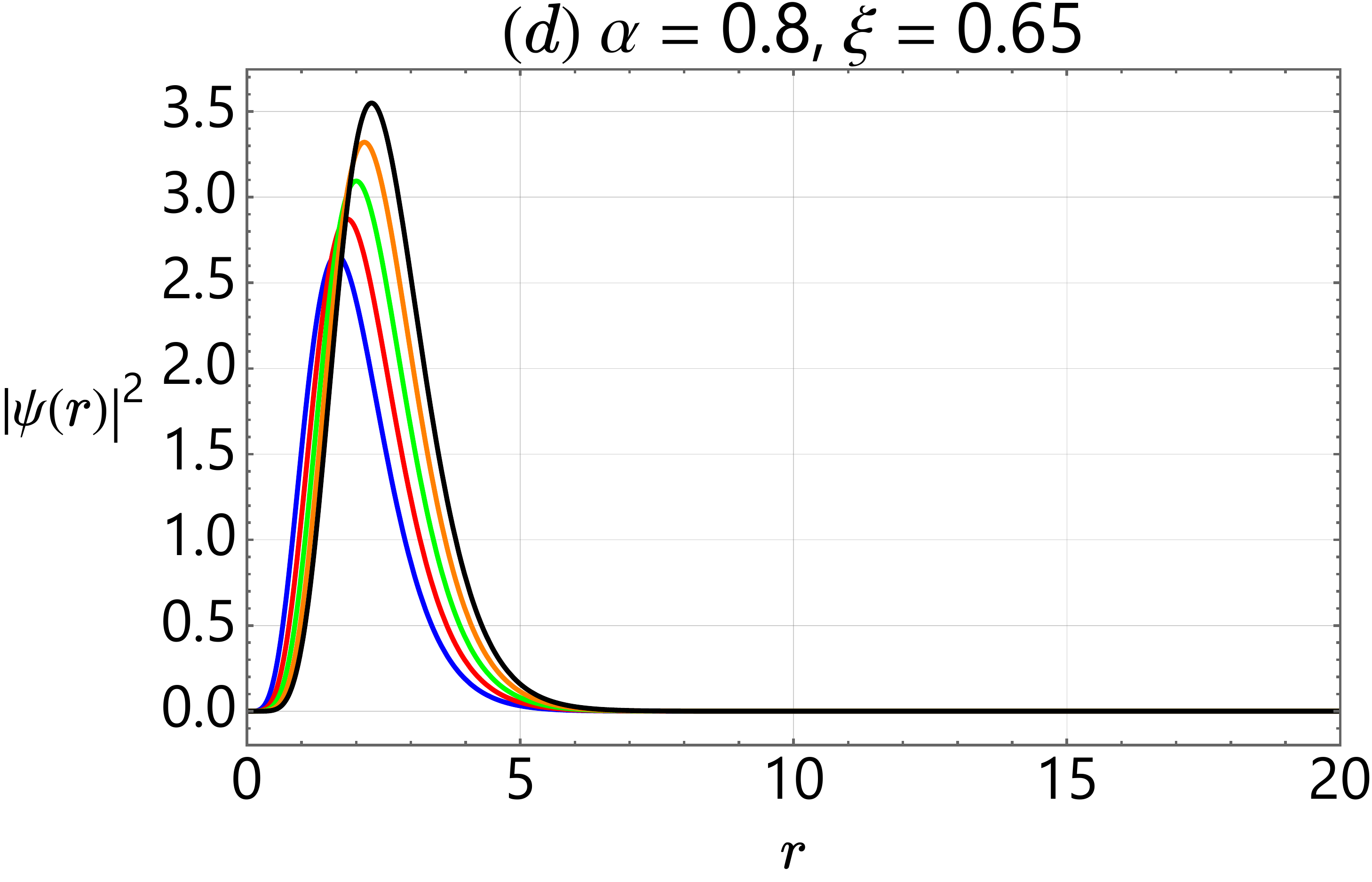}
	\caption{The plots of $|u(r)|^2$ as a function of $r$ for different values of $n$ displayed for (a) $\alpha=0.2$, $\xi=01$, (b) $\alpha=0.8$, $\xi=0.1$, (c) $\alpha=0.2$, $\xi=0.65$, and (d) $\alpha=0.8$, $\xi=0.65$. We use the parameters $\hbar = 1$, $M=1$, $e=1$, $k=1$, and $l=1$.}
	\label{Fig_DP1}
\end{figure}
\begin{figure}[!h]
	\centering
	\includegraphics[scale=0.25
	]{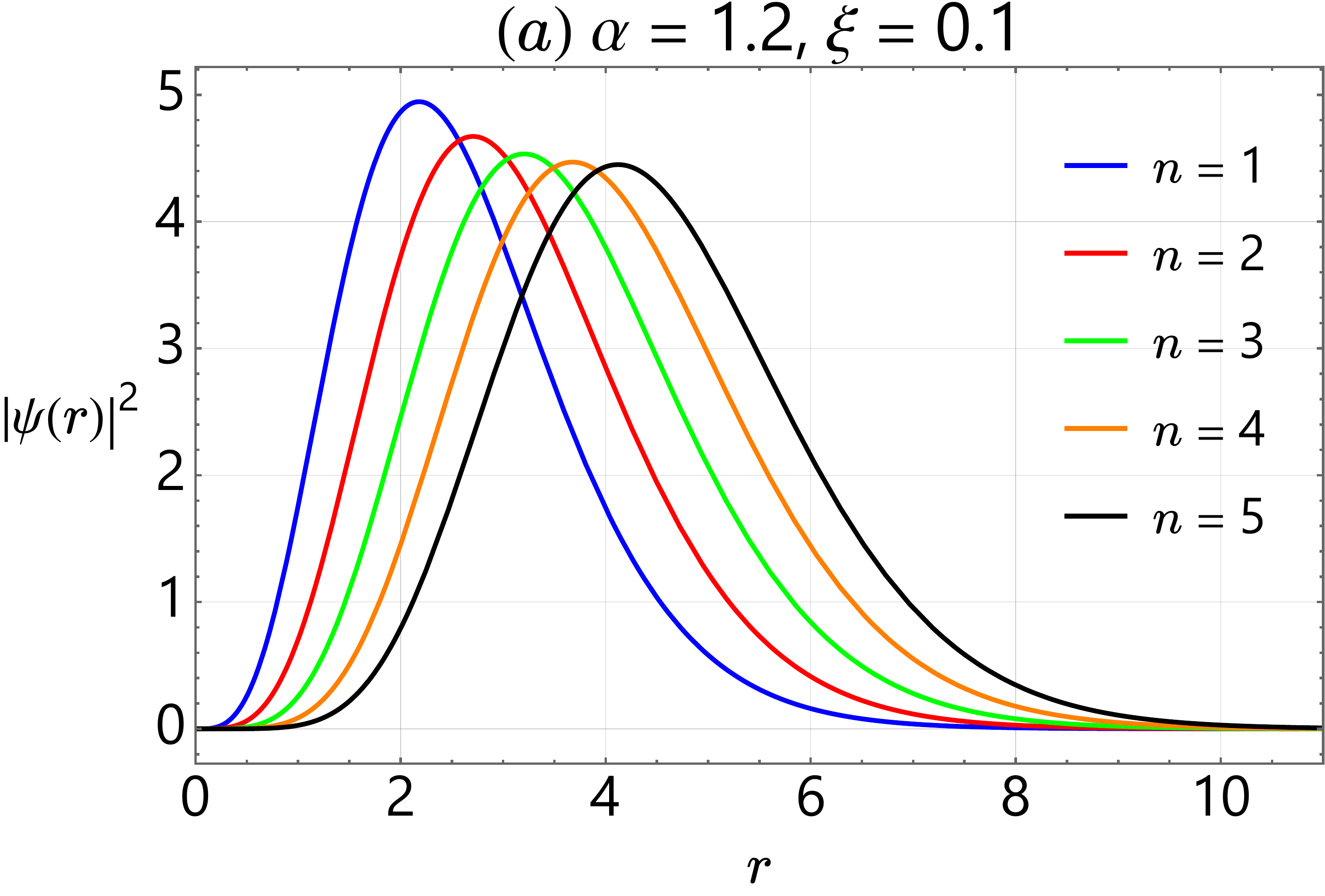}\qquad
	\includegraphics[scale=0.25
	]{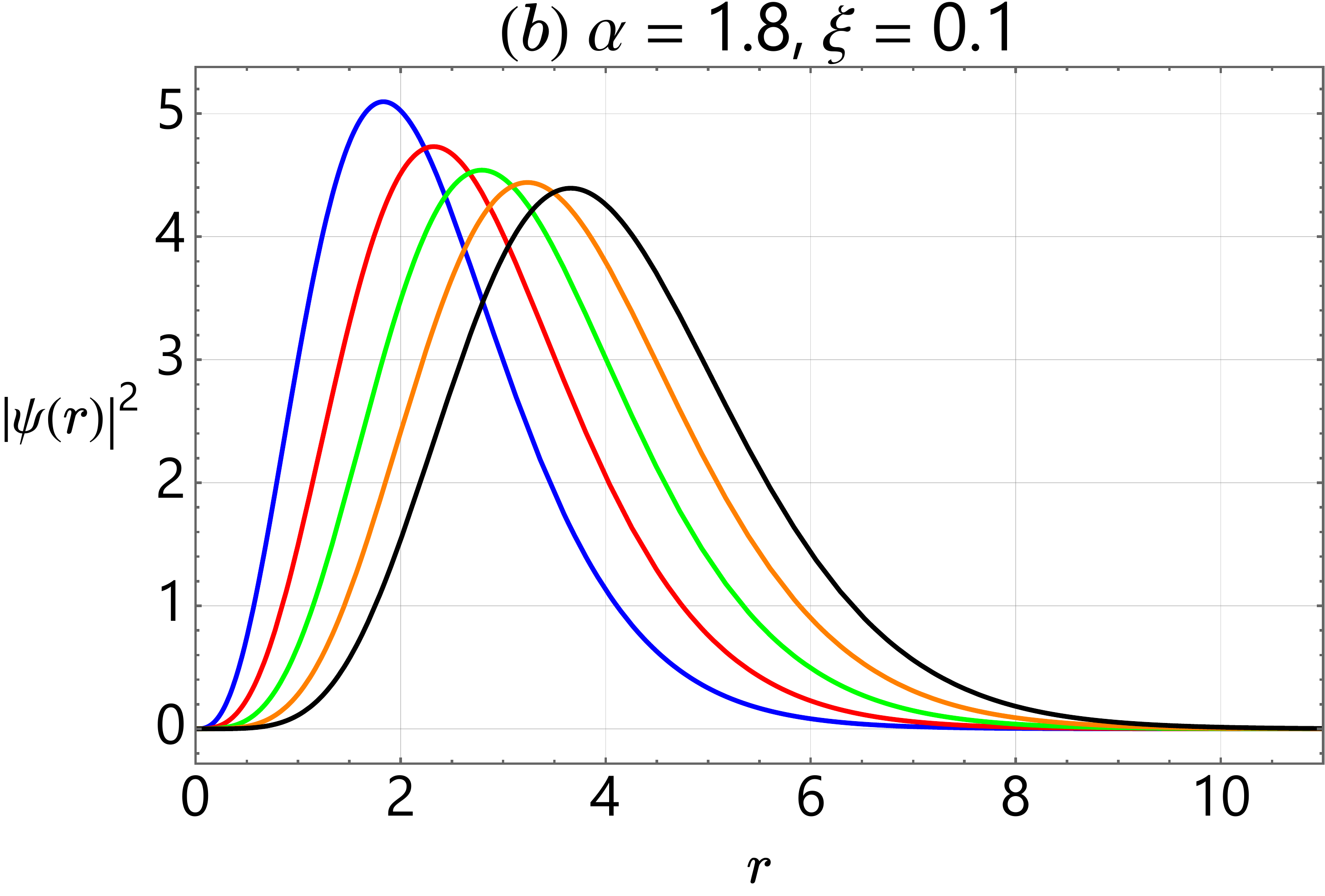}
	\includegraphics[scale=0.25
	]{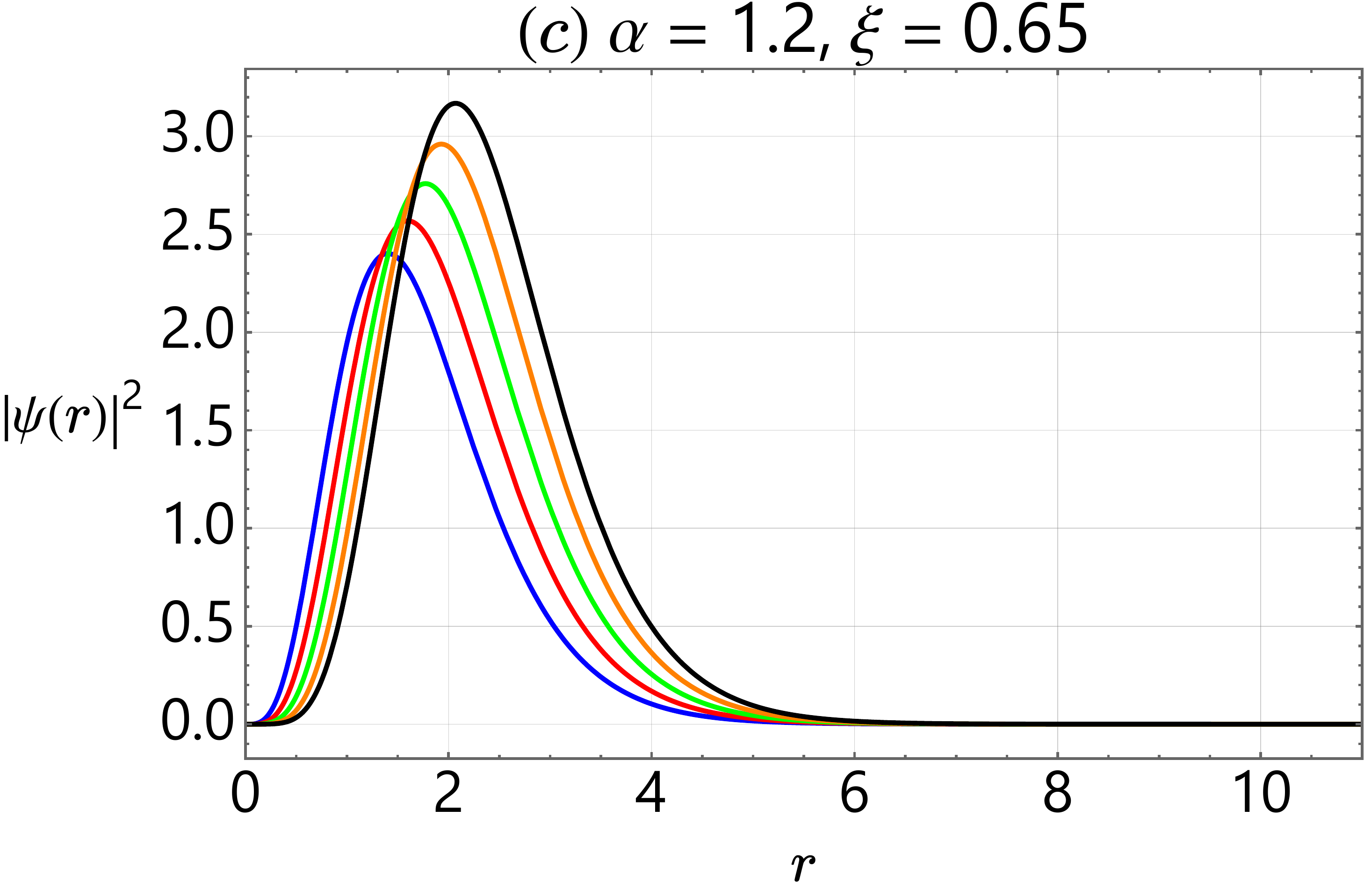}\qquad
	\includegraphics[scale=0.25
	]{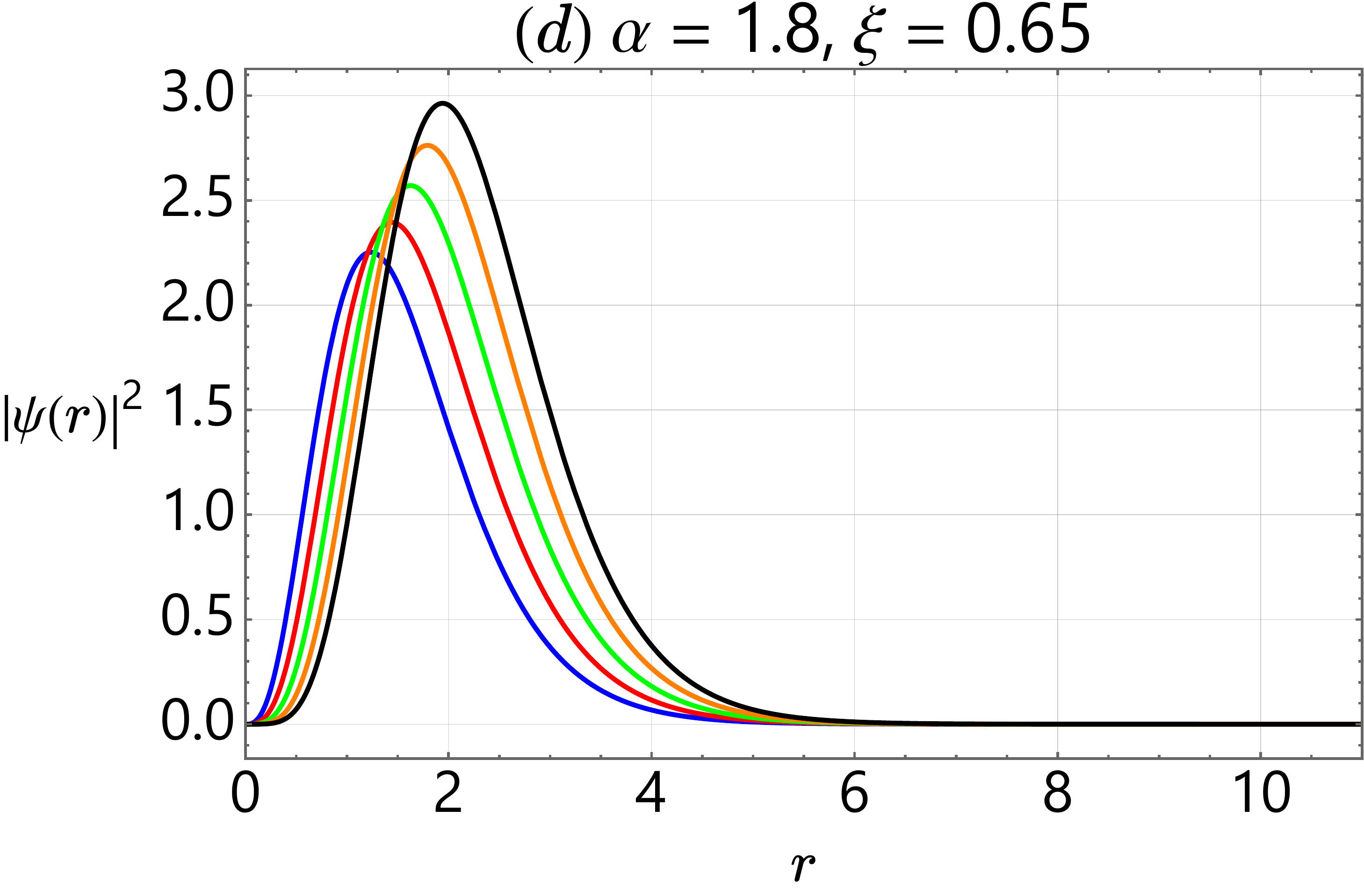}\qquad
	\caption{The plots of $|u(r)|^2$ as a function of $r$ for different values of $n$ displayed for (a) $\alpha=1.2$, $\xi=0.1$, (b) $\alpha=1.8$, $\xi=0.1$, (c) $\alpha=1.2$, $\xi=0.65$, and (d) $\alpha=1.8$, $\xi=0.65$. We use the parameters $\hbar = 1$, $M=1$, $e=1$, $k=1$, and $l=1$.}
	\label{Fig_DP2}
\end{figure}
In Equation (\ref{gs}), the function $(1-y)^{-i\kappa}$ represents an arbitrary choice of the particular conditions for a scattering wave, namely the outgoing wave at
infinity (see Equation (\ref{bcc})).
In Figures \ref{Fig_DP1} and \ref{Fig_DP2}, we make plots of  $|u(y)|^2$ for different values of $n$. In both plots, we use $\hbar = 1$, $M=1$, $e=1$, $k=1$, and $l=1$. For the values of $\xi$, we use some values based on others in the literature \cite{CPB.2009.18.3663,PLA.2019.383.3010,CPB.2018.27.020301,PLA.2008.372.4779}. For $\xi=0.1$ and $\alpha=0.2$, the amplitudes of $|u(r)|^{2}$ increase as $n$ are increased (see the solid blue line for $n=1$ and the solid black line for $n=5$, respectively) (Figure \ref{Fig_DP1}(a)). Keeping $\xi$ at $0.1$ and increasing $\alpha$ to $0.8$, we observe the reverse effect, i.e., the curves with the largest amplitude are those with an increasingly smaller $n$ (see the blue solid line for $n=1$). Furthermore, $|u(r)|^{2}$ becomes more localized, with the maximum of the amplitudes moving to the left (Figure \ref{Fig_DP1}(b)). When we analyze the profile of $|u(r)|^{2}$ for $\alpha=0.65$ and keep the other parameters, we see that the amplitudes of $|u(r)|^{2}$ for $\alpha=0.2$ are larger than those obtained for the case with $\alpha=0.8$. In both cases, the amplitudes increase when $n$ is increased. It is important to analyze the profiles in Figure \ref{Fig_DP1} for $\alpha>1$ (Figure \ref{Fig_DP2}). For the respective values $\alpha=1.2$, $\xi=0.1$ and $\alpha=1.8$, $\xi=0.1$, only a reduction in the amplitude of $|u(r)|^{2}$ occurs when the parameter $\alpha$ is increased. In both cases, the amplitudes decrease when $n$ increases (Figures \ref{Fig_DP2}(a)-(b)). As in Figure \ref{Fig_DP1}(a)-(b), when we increase $\xi$ to $0.65$ ((Figures \ref{Fig_DP2}(c)-(d)), respectively) and keep the same values of $\alpha$, only a small variation in the amplitude of $|u(r)|^{2}$ is observed. The increase in amplitude occurs when the quantum number $n$ is increased. In the order in which they are displayed, the amplitude with $n=1$ (curve with blue solid line) is the smallest, while the amplitude with $n=5$ (curve with black solid line) is the largest.

To access other properties of the hypergeometric function $_{2}	{F}_{1}\left( a,b;\,c;\,y\right)$ in Equation (\ref{gs}), the reader is invited to see References \cite{aomoto2011theory,abramo}. Starting from the solution (\ref{gs}), we can now study different properties of the system, such as the scattering phase shift, the scattering matrix ($S$-matrix), and the energies for bound states. We address these topics in the next sections. 

\section{Scattering phase shift}\label{sec3}

In this section, we shall derive the phase shift of the wave functions corresponding to the solution (\ref{gs}). For this purpose, we consider the boundary condition such that when $y\rightarrow 0$ ($r\rightarrow 0$), $u\left( y\right) $, it is finite. Solution (\ref{gs}) satisfies these requirements. Thus, returning to the variable $r$, the solution becomes
\begin{equation}
	u\left( r\right) =\mathit{c}_{n}\,\left( 1-e^{-\xi r}\right) ^{d}e^{ikr}\,_{2}{F}_{1}\left( a,b;\,c;\,1-e^{-\xi r}\right) ,  \label{rels}
\end{equation}
where $\mathit{c}_{n}$ is the normalization constant. To continue solving the problem, we need to associate the solution $_{2}{F}_{1}\left( a,b;\,c;\,1-e^{-\xi r}\right)$ to some transformation relation allowing us to analyze its asymptotic behavior. Such a transformation is obtained through the analytical continuation of the hypergeometric function given by \cite{Book.2010.NIST}
\begin{align}
	_{2}{F}_{1}\left( a,b;c;y\right) & =\frac{\Gamma \left( c\right) \Gamma
		\left( c-a-b\right) }{\Gamma \left( c-a\right) \Gamma \left( c-b\right) }\, _{2}{F}_{1}\left( a,b;a+b-c+1;1-y\right)  \notag \\
	& +\left( 1-y\right) ^{c-a-b}\frac{\Gamma \left( c\right) \Gamma \left(
		a+b-c\right) }{\Gamma \left( a\right) \Gamma \left( b\right) }\,_{2}{F}_{1}\left( c-a,c-b;c-a-b+1;1-y\right)\label{fg}
\end{align}
together with the results
\begin{align}
	&_{2}{F}_{1}\left( a,b;a+b-c+1;0\right) =1,  \label{rel1}\\
	&_{2}{F}_{1}\left( c-a,c-b;c-a-b+1;0\right) =1. \label{rel2}
\end{align}
Then, using (\ref{rel1}) and (\ref{rel2}) in Equation (\ref{fg}), the desired transformation it is found to be \cite{CPB.2009.18.3663}
\begin{equation}
	_{2}{F}_{1}\left( a,b;c;1-e^{-\xi r}\right) \rightarrow \frac{\Gamma \left( c\right)
		\Gamma \left( c-a-b\right) }{\Gamma \left( c-a\right) \Gamma \left(c-b\right) }{~}+e^{-\xi(c-a-b) r}\,\frac{\Gamma \left( c\right)\Gamma \left( a+b-c\right) }{\Gamma \left( a\right) \Gamma \left( b\right) }.\label{lim}
\end{equation}
We aim to write Equation (\ref{lim}) in a form that allows us to find an expression for the phase shift. First, the quantities $c-a-b$ and $a+b-c$ in Equation (\ref{lim}) are given by $2i\kappa$
and $-2i\kappa$, respectively, from which we can verify that
\begin{equation}
	a+b-c=\left(c-a-b\right)^{\ast}.\label{d1}
\end{equation}
Similarly, we can also show that 
\begin{align}
	c-a& =d+i\kappa -\Delta =b^{\ast },  \label{d2} \\
	c-b& =d+i\kappa +\Delta =a^{\ast }.  \label{d3}
\end{align}
Using these results, Equation (\ref{lim}) can be written as
\begin{equation}
	_{2}{F}_{1}\left( a,b;c;y\right) \rightarrow \Gamma \left( c\right) \left[ 
	\frac{\Gamma \left(c-a-b\right)}{\Gamma \left( c-a\right) \Gamma \left(
		c-b\right) }{~}+e^{-2ikr}\frac{\Gamma \left( c-a-b\right) ^{\ast}}{\Gamma \left( c-b\right) ^{\ast }\Gamma \left( c-a\right) ^{\ast }}\right]. \label{rf}
\end{equation}
Taking into account the relations
\begin{equation}
	\frac{\Gamma \left( c-a-b\right) }{\Gamma \left( c-a\right) \Gamma \left(
		c-b\right) }=\left\vert \frac{\Gamma \left( c-a-b\right) }{\Gamma \left(
		c-a\right) \Gamma \left( c-b\right) }\right\vert e^{i\delta_l },
\end{equation}
\begin{equation}
	\frac{\Gamma \left( c-a-b\right) ^{\ast }}{\Gamma \left( c-a\right) ^{\ast
		}\Gamma \left( c-b\right) ^{\ast }}=\left\vert \frac{\Gamma \left(
		c-a-b\right) }{\Gamma \left( c-a\right) \Gamma \left( c-b\right) }
	\right\vert e^{-i\delta_l },
\end{equation}
and substituting them into Equation (\ref{rf}), we find
\begin{equation}
	_{2}{F}_{1}\left( a,b;c;y\right) \rightarrow \Gamma \left( c\right)
	\left\vert \frac{\Gamma \left( c-a-b\right) }{\Gamma \left( c-a\right)
		\Gamma \left( c-b\right) }\right\vert e^{-ikr}\left[ e^{i\left(kr+\delta_l \right) }{~}+e^{-i\left( kr+\delta_l \right) }\right].  \label{f2}
\end{equation}
This equation can be written in a more convenient form. First, we use the identity $2\cos x =e^{ix }+e^{-ix}$ and then rewrite $2\cos
\left( kr+\delta_l \right) =e^{i\left(kr+\delta_l \right) }+e^{-i\left(kr+\delta_l \right) }$. Next, by certifying that $\cos \left( kr+\delta_l \right) =\sin \left( kr+\pi /2+\delta_l
\right) $, we make
\begin{equation}
	\cos \left( kr+\theta \right) =\sin \left( kr-\frac{\pi \ell
	}{2}+\frac{\pi }{2}\left( \ell+1\right) +\delta_l \right).  \label{nr}
\end{equation}
The expression above provides us with the asymptotic behavior of the solution $
u\left( y\right) $ for $r\rightarrow \infty $, i.e.
\begin{equation}
	u\left( y\right) \sim \sin \left( kr-\frac{\pi \ell}{2}+\frac{\pi }{2}\left(
	\ell+1\right) +\delta_l \right).
\end{equation}
Comparing this result with the boundary condition (\ref{bc}), the phase shift $\delta_\theta$ is found, and its expression is given by
\begin{equation}
	\delta_l =\frac{\pi }{2}\left( \ell+1\right) +\arg \Gamma \left( c-a-b\right)
	-\arg \Gamma \left( c-a\right) -\arg \Gamma \left( c-b\right).  \label{phase}
\end{equation}
Substituting the parameters  (\ref{d1}), (\ref{d2}) and (\ref{d3}) into (\ref
{phase}), we obtain
\begin{equation}
	\delta_l =\frac{\pi }{2}\left( \ell+1\right) +\arg \Gamma \left( 2i\kappa \right)
	-\arg \Gamma \left( d+i\kappa -\Delta \right) -\arg \Gamma \left( d+i\kappa
	+\Delta \right).  \label{spr}
\end{equation}
It is important to emphasize that the phase shift depends explicitly on the parameters $\alpha$ and $\xi$, through $\ell$, $d$, and $\Delta$, which shows that $\delta_{l}$ is affected by the curvature generated by the global monopole.

\section{Analysis of bound states}\label{sec4}

Similarly to the Coulomb potential, the Hulth\'{e}n potential also admits bound state solutions. The bound state energies can be found from the $S$-matrix. According to general scattering theory, the poles of the $S$-matrix in the upper half of the complex plane are associated with the bound state energies. Using the result for the phase shift (\ref{spr}), the $S$-matrix can be written as 
\begin{align}
	S &=e^{2i\delta_l}, \\
	&=e^{i\pi \left( \ell+1\right) }e^{2i\arg \Gamma \left( 2i\kappa \right)
	}e^{-2i\arg \Gamma \left( d+i\kappa -\Delta \right) }e^{-2i\arg \Gamma
		\left( d+i\kappa +\Delta \right)}.
\end{align}
The poles of the S-matrix are given by the poles of the gamma functions $\Gamma \left( d+i\kappa -\Delta \right) $ and $\Gamma \left( d+i\kappa +\Delta\right) $ in Equation (\ref{spr}). However, we must remember that the function $\Gamma \left( z\right) $ has poles at $z=-n$, where $n$ is a non-negative integer. Thus, analyzing the poles of $\Gamma \left( d+i\kappa
-\Delta \right) $, we get
\begin{equation}
	d+i\kappa -\Delta =-n.
\end{equation}
Using the parameters given in Equation (\ref{rd}), we find
\begin{equation}
	d+i\frac{\sqrt{2ME_{nl}}}{\hbar \alpha \xi }-\sqrt{\wp ^{2}-\frac{2ME_{nl}}{\hbar
			^{2}\alpha ^{2}\xi ^{2}}}=-n. \label{rlE}
\end{equation}
Finally, solving Equation (\ref{rlE}) for $E_{nl}$, we obtain the energy eigenvalues
\begin{equation}
	E_{nl}=-\frac{(d+n-\mathcal{\wp })^{2}(d+n+\mathcal{\wp })^{2}}{\beta (d+n)^{2}},%
	\text{ with }\beta =\frac{8M}{\alpha ^{2}\hbar ^{2}\xi ^{2}}.\label{sve}
\end{equation}
These energies can also be obtained by solving Equation (\ref{edoH}) for bound states. We make this by solving equation (\ref{edoH}) via the Frobenius method. We use solutions of the form
\begin{equation}
	u\left( y\right) =y^{\gamma }\left( 1-y\right) ^{\nu }h\left( y\right)
	\label{t1},
\end{equation}
where $\nu$ and $\gamma$ are arbitrary constants to
be determined and where $h(y)$ is an unknown function. Note that the solution (\ref{t1}) is finite at regular singular points $y=0$, $y=1$, and $y=\infty$. Substituting this solution into equation (\ref{edoH}), we find the differential equation
\begin{align}
	h^{\prime \prime }\left( y\right) +&\left[ \frac{2\gamma -\left( 1+2\gamma
		+2\nu \right) y}{y\left( 1-y\right) }\right] h^{\prime }\left( y\right) +
	\frac{\nu ^{2}-\kappa _{b}^{2}}{\left( 1-y\right) ^{2}}h\left( y\right) 
	\notag \\
	& +\left[ \frac{\wp ^{2}-2\gamma \nu -\gamma ^{2}}{y\left( 1-y\right) }
	\right] h\left( y\right) +\frac{\gamma \left( \gamma -1\right) -\lambda ^{2}
	}{y^{2}\left( 1-y\right) }h\left( y\right) =0, \label{hf}
\end{align}
where $\kappa_{b}=k_{b}/\xi$, $\ k_{b}=\sqrt{-2ME_{nl}}/\hbar \alpha>0$, with $``b"$ labeling bound states.
The parameters $\nu$ and $\gamma$ are determined by canceling out the coefficients 
\begin{align}
	&\nu ^{2}-\kappa _{b}^{2}=0, \label{c01}\\
	&\gamma \left( \gamma -1\right) -\lambda ^{2}=0, \label{c02}
\end{align}
from which we find
\begin{equation}
	\nu _{1} =+\kappa _{b} \;\text{or}\; \nu _{2} =-\kappa _{b} \label{cd1}
\end{equation}
and
\begin{equation}
	\gamma _{1} =d\; \text{or}\;
	\gamma _{2} =1-d,
\end{equation}
respectively. For bound state solutions, we must choose $\nu _{1}$ and $\gamma _{1}$ above, which leads to the equation
\begin{equation}
	y\left( 1-y\right) h^{\prime \prime }\left( y\right) +\left[ \zeta
	_{3}-\left( 1+\zeta _{1}+\zeta _{2}\right) y\right] h^{\prime }\left( y\right)
	-\zeta _{1}\zeta _{2}h\left( y\right) =0,  \label{ehp}
\end{equation}
where
\begin{align}
	\zeta _{1} &=\gamma _{1}+\nu _{1}+\sqrt{\wp ^{2}+\nu _{1}^{2}}, \\
	\zeta _{2} &=\gamma _{1}+\nu _{1}-\sqrt{\wp ^{2}+\nu _{1}^{2}},\\
	\zeta _{3} &= 2\gamma _{1}.
\end{align}
Equation (\ref{ehp}) is a hypergeometric differential equation. It is known that the singular points of this equation are regular. Therefore, we can assume series solutions around $y=0$ of the form
\begin{equation}
	h\left( y\right) =\sum\limits_{s=0}^{\infty }a_{s}\,y^{s+c},\;\text{with}\;
	a_{0}\neq 0.\label{sh}
\end{equation}
By substituting the solution (\ref{sh}) into the differential equation (\ref{ehp}), we obtain the indicial equation 
\begin{equation}
	a_{0}\left[ c\left( c-1\right) +\zeta_{3} c\right] =0,
\end{equation}
whose roots are 
\begin{align}
	c_{1} &=0, \\
	c_{2} &=1-\zeta_{3} .
\end{align}
and the recurrence relation
\begin{equation}
	a_{s+1}=\frac{\left( s+c\right) \left( s+c+\zeta_{1} +\zeta_{2} \right) +\zeta_{1}
		\zeta_{2}}{\left( s+c+1\right) \left( s+c+\zeta_{3} \right) }a_{s},\;\text{for}\;
	s\geqslant 0.  \label{rr}
\end{equation}
The general solution of (\ref{ehp}) is written as
\begin{equation}
	h\left( y\right) =A\,h_{1}\left( y\right) +B\,h_{2}\left( y\right),\label{slut}
\end{equation}
where $A$ and $B$ are, respectively, the coefficients of the regular and irregular solutions at the origin, with
\begin{equation}
	h_{1}\left( y\right) =\sum\limits_{s=0}^{\infty }\frac{\left( \zeta_{1}
		\right) _{s}\left( \zeta_{2} \right) _{s}}{\left( 1\right) _{s}\left( \zeta_{3}
		\right) _{s}}\,y^{s}=\,_{2}F_{1}\left( \zeta_{1} ,\zeta_{2} ,\zeta_{3} ;y\right)
	\label{sc1}
\end{equation} 
and
\begin{align}
	h_{2}\left( y\right) &=\,y^{1-\zeta_{3} }\sum\limits_{s=0}^{\infty }\frac{\left( \zeta_{1}  +1-\zeta_{3}  \right)_{s}\left(\zeta_{2}  +1-\zeta_{3}  \right)_{s}}{
		\left(2-\zeta_{3}  \right) _{s}\left( 1\right) _{s}}\,y^{s},\\
	&=\,y^{1-\zeta_{3}}~_{2}F_{1}\left( \zeta_{1} +1-\zeta_{3}
	,\zeta_{2} +1-\zeta_{3} ,2-\zeta_{3} ;y\right).
\end{align}
The solution (\ref{t1}) is given by
\begin{align}
	u\left( y\right) &=A\,y^{d}\left( 1-y\right) ^{\kappa _{b}}{}_{2}F_{1}\left(
	\zeta _{1},\zeta _{2},2d;y\right) \notag\\&+B\,y^{1-d}\left( 1-y\right) ^{\kappa
		_{b}}~_{2}F_{1}\left( \zeta_{1}+1-2d,\zeta _{2}+1-2d,2-2d;y\right) 
\end{align}
Since $\zeta_{3}$ is not an integer and $u(y)=0$ at $y=0$ (or $r \rightarrow \infty$), we shall take $B=0$. Thus, the relevant solution is 
\begin{align}
	u\left( y\right) &=C_{nl}\,y^{d}\left( 1-y\right) ^{\kappa _{b}}{}_{2}F_{1}\left(
	\zeta_{1},\zeta _{2},2d;y\right), \label{gsol}
\end{align}
where $C_{nl}$ is the normalization constant.
To obtain bound states energies, we must require that the series $_{2}F_{1}\left(
\zeta_{1},\zeta _{2},2d;y\right)$ terminates, resulting in a polynomial of degree $n$. This means that in the recurrence relation (\ref{rr}), we must impose that $a_{n+1}=0$, which leads to (with $c_{1}=0$)
\begin{equation}
	n\left( n+\zeta_{1} +\zeta_{2} \right) +\zeta_{1} \zeta_{2} =0  \label{rel}
\end{equation}
or
\begin{equation}
	2n\kappa_{b} +d^{2}-\wp ^{2}+2d\kappa_{b} +2dn+n^{2}=0,
\end{equation}
which solved for $\kappa_{b}$ provides
\begin{equation}
	\kappa_{b} =\frac{\wp ^{2}-\left( d+n\right) ^{2}}{2\left( n+d\right) }>0,\;\text{with}\;n+d \neq 0.  \label{Er}
\end{equation}
Using the relation
\begin{equation}
	\kappa_{b}=\sqrt{-\frac{2ME_{nl}}{\hbar ^{2}\alpha ^{2}\xi ^{2}}}>0,\label{cdk}
\end{equation}
in Equation (\ref{Er}), and solving the resulting equation for $E_{nl}$, we obtain
\begin{equation}
	E_{nl}=-\frac{(d+n-\mathcal{\wp })^{2}(d+n+\mathcal{\wp })^{2}}{\beta (d+n)^{2}},
	\text{ with }\beta =\frac{8M}{\alpha ^{2}\hbar ^{2}\xi ^{2}},\label{energy}
\end{equation}
which is just the Equation (\ref{sve}). To ensure the validity of the relation (\ref{cdk}), we must require in Equation (\ref{Er}) that
\begin{equation}
	\wp ^{2}>\left( d+n\right) ^{2},  \label{enq}
\end{equation}
which gives the upper values for $n$
\begin{align}
	n& <\left\vert \wp \right\vert -d,\;\text{with}\;d\in \mathbb{R}\;\;\wp \in 
	\mathbb{R}
\end{align}
or more explicitly
\begin{equation}
	n <\left\vert \frac{2M}{\hbar ^{2}\alpha ^{2}\xi}\left( Ze^{2}-\mathcal{K}\left( \alpha \right)\right)
	\right\vert -\left[ \frac{1}{2}+{\frac{1}{2}\sqrt{1+\,\frac{4l\left(
				l+1\right) }{\alpha ^{2}}}}\right].\label{cvl}
\end{equation}
Equation (\ref{cvl}) establishes a condition for the occurrence of bound states and, in addition, determines the range for the quantum number $n$ where such states must appear. It should be emphasized that the energies (\ref{energy}) could be obtained directly by substituting $\kappa \rightarrow ik_{b}$ in Equation (\ref{gs}). This leads us to the solution (\ref{gsol}). The characteristics of the hypergeometric function are well known. As mentioned above, when $\zeta_{1}=-n$, with $n=0,1,2,\ldots,$ and $\zeta_{3}\neq 0,-1,-2,\ldots ,$ the function $ _{2}F_{1}(\zeta_{1},\zeta_{2},\zeta_{3};y)$ becomes a polynomial. In this way, the wave function for bound states (as a function of $r$) reads 
\begin{equation} 
	u\left( r\right) =\mathit{C}_{nl}\,\left( 1-e^{-\xi r}\right) ^{d}e^{-k_{b}r}\,_{2}{F}_{1}\left( -n,d\,+\,\kappa_{b}-\sqrt{\wp ^{2}+\kappa_{b}^{2}},\,2d;\,1-e^{-\xi r}\right) ,
	\label{sr}
\end{equation}
and the expression for the  bound state energies can be found in the condition  
\begin{equation}
	d\,+\,\kappa_{b}+\sqrt{\wp ^{2}+\kappa_{b}^{2}} =-n,  \label{cnd}
\end{equation}
which solved for $E_{nl}$ leads us again to Equation (\ref{sve}), which written in its explicit form reads
\begin{equation} 
	E_{nl}=-\frac{\alpha ^{2}\hbar ^{2}\xi ^{2}}{8M}\left[ \frac{\frac{2M\xi }{\hbar
			^{2}\alpha ^{2}\xi ^{2}}\left( Ze^{2}-\mathcal{K}\left( \alpha \right)
		\right) }{n+\frac{\sqrt{4l\left( l+1\right) +\alpha ^{2}}}{2\alpha }+\frac{1
		}{2}}-\left( n+\frac{\sqrt{4l\left( l+1\right) +\alpha ^{2}}}{2\alpha }+
	\frac{1}{2}\right) \right] ^{2}.\label{enexp}
\end{equation}
It can be verified that for $\xi \rightarrow 0$ and keeping the other parameters fixed in Equation (\ref{enexp}), $E_{nl}$ assumes finite values. In Table \ref{tb1}, we show some energy values for $\alpha=0.7$ and different values of $l$.
\begin{table}[!h]
	\caption{Energies in the limit $\xi \rightarrow 0$ for $n=1$ and different values of $l$ (Equation (\ref{enexp})). We assume $\hbar =1, M=1, Z=1, e=1, n=1$ and $\alpha =0.7$.}
	\label{tb1}
	\begin{center}
		\begin{tabular}{clr}
			\hline
			&$E_{n,l}$    &Values\\
			\hline \hline
			&$E_{1,1}$    &$-0.0472842$\\
			&$E_{1,2}$    &$-0.0236029$\\
			&$E_{1,3}$    &$-0.0141781$\\
			&$E_{1,4}$    &$-0.0094620$\\
			&$E_{1,5}$    &$-0.0067639$\\
			&$E_{1,6}$    &$-0.0050758$\\
			&$E_{1,7}$    &$-0.0039496$\\
			&$E_{1,8}$    &$-0.0031608$\\
			&$E_{1,9}$    &$-0.0025868$\\
			&$E_{1,10}$   &$-0.0021561$\\
			\hline
		\end{tabular}
	\end{center}
\end{table}
It is important to mention that for $Ze^{2}=0$ (absence of the Hulthén potential), the system admits bound states only for $\mathcal{K}(\alpha)<0$, i.e., only for the attractive electrostatic self-
interaction. However, with the presence of the Hulthén potential, the system also admits bound states for $\mathcal{K}(\alpha)>0$. Indeed, this is evidenced in Figures \ref{Fig_veff} and \ref{Fig_veff2}, where the result of the superposition between the potentials $V_{H}\left( r\right)$ and $V_{SI}\left( r\right)$ is shown more explicitly. Therefore, bound states are possible only when $V_{H}\left( r\right)$ + $V_{SI}\left( r\right)<0$.
\begin{figure}[!h]
	\centering
	\includegraphics[scale=0.26]{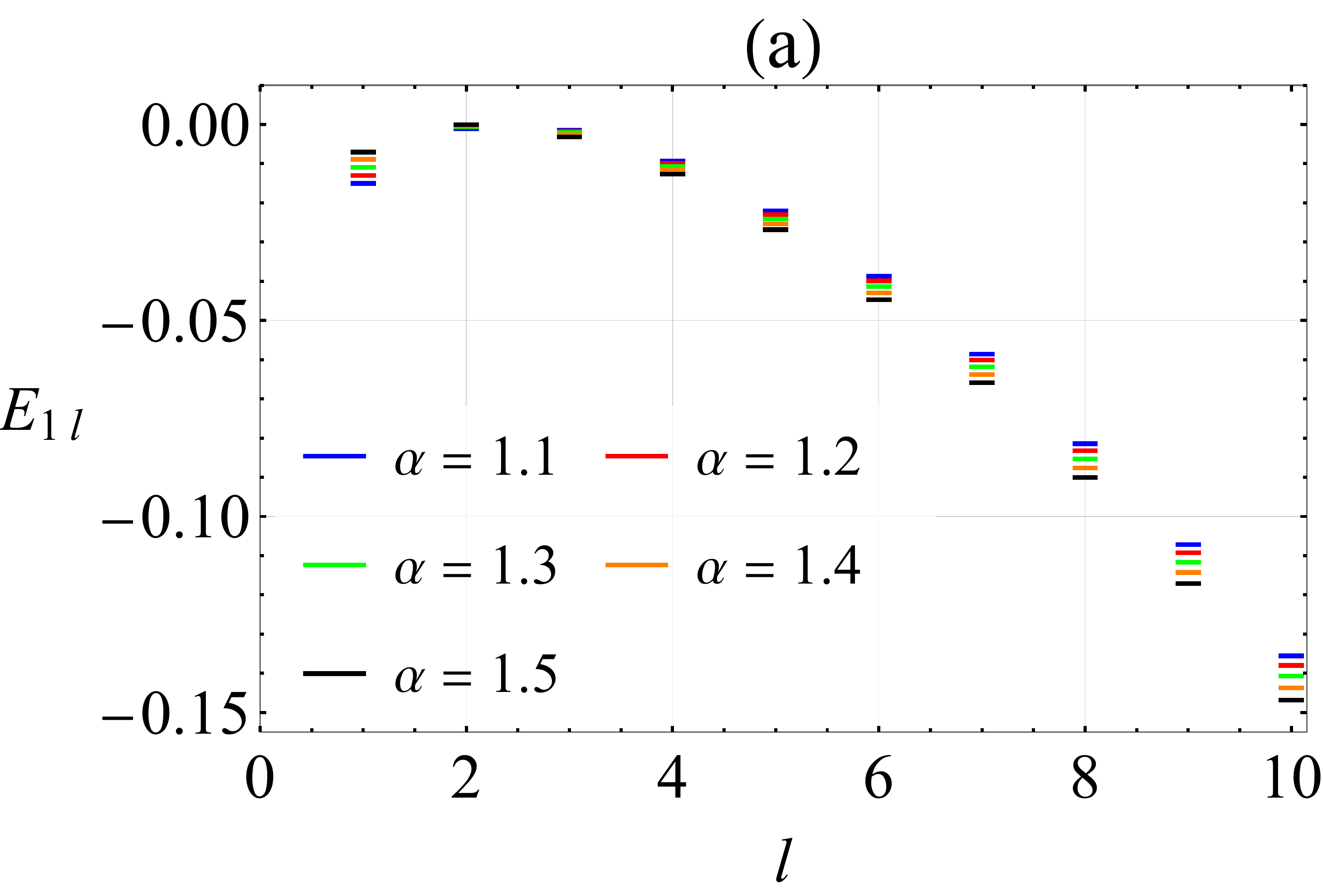}
	\includegraphics[scale=0.26]{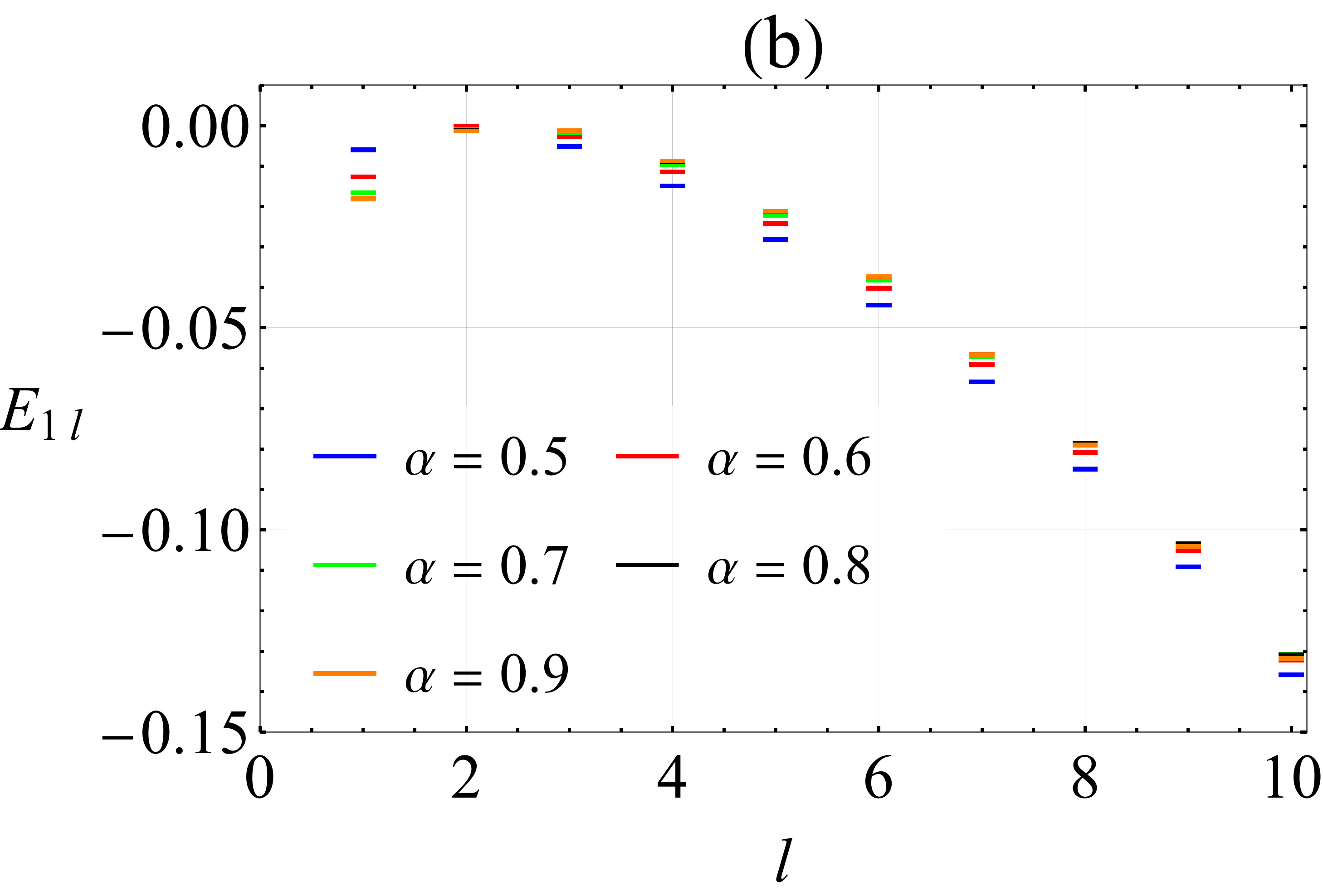}
	\caption{Energy levels (Equation (\ref{enexp})) with $n=1$ as a function of $l$ for $\xi=0.1$. In (a), we display the energies corresponding to different values of $\alpha>1$ and in (b) for values of $\alpha <1$. An inversion between the two profiles is observed at $l=2$.}
	\label{Energy_l}
\end{figure}
We can study the energies (\ref{enexp}) by sketching them as a function of the parameters involved. In all the energy plots that we illustrate here, we choose for convenience to analyze the state with $n=1$ and use $\hbar=1$, $M=1$, $Z=1$, and $e=1$. In Figure \ref{Energy_l}, we plot the energy levels with $n=1$ for different values of $l$. We can see that $|E_{10}|>|E_{11}|>|E_{12}|$, following the order of the $\alpha$ values considered. We also see that the separation between the energy levels with $l=0$ corresponding to a given value of $\alpha$ is greater than the separation between the energy levels with $l>0$. For $l>2$, we observe an inversion between the energy levels with a given value of $\alpha$ when we compare them with the energy levels with $l<2$. In this case, $|E_{1l}|$ increases for $l>2$.
\begin{figure}[!h]
	\centering
	\includegraphics[scale=0.25
	]{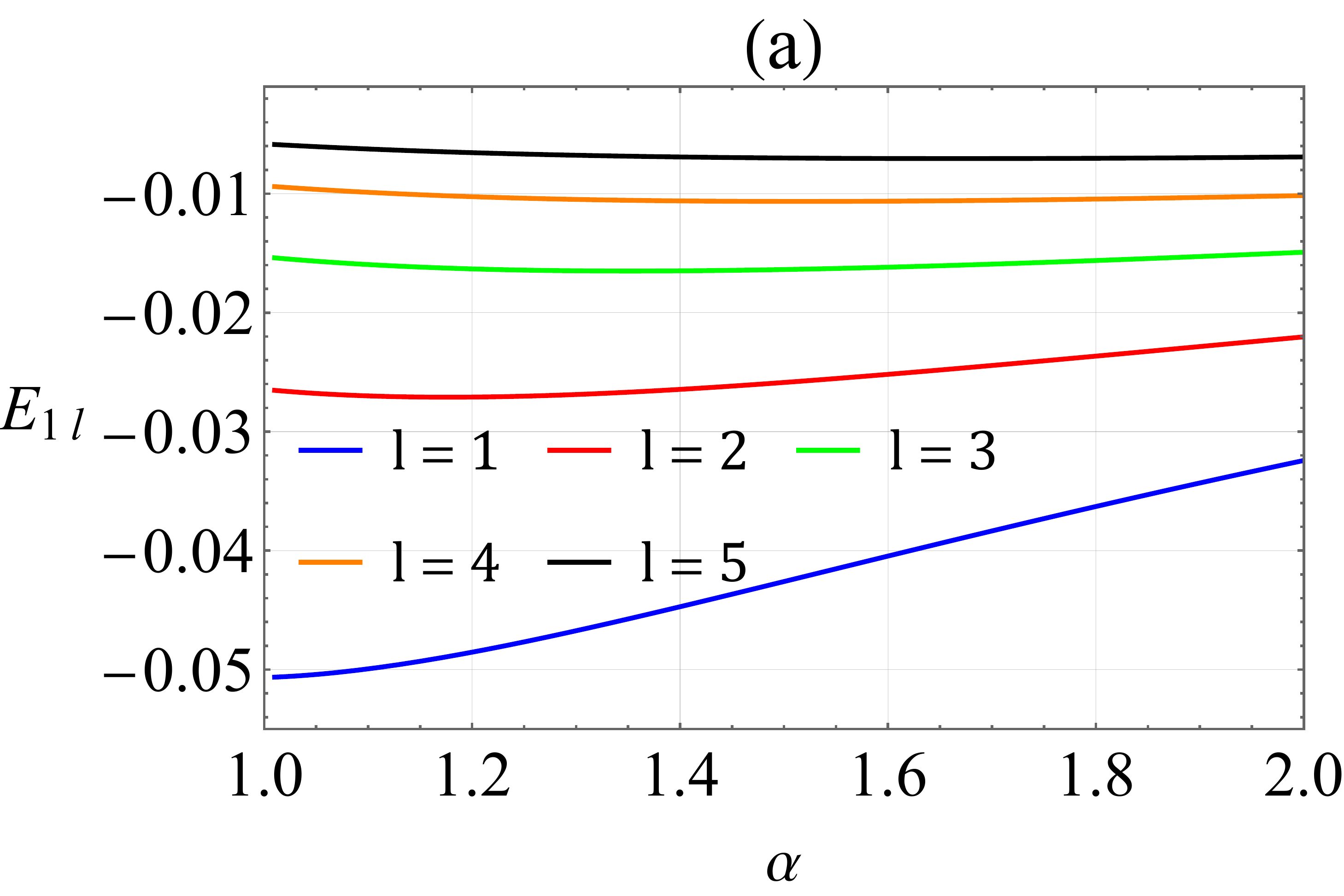}\qquad
	\includegraphics[scale=0.25
	]{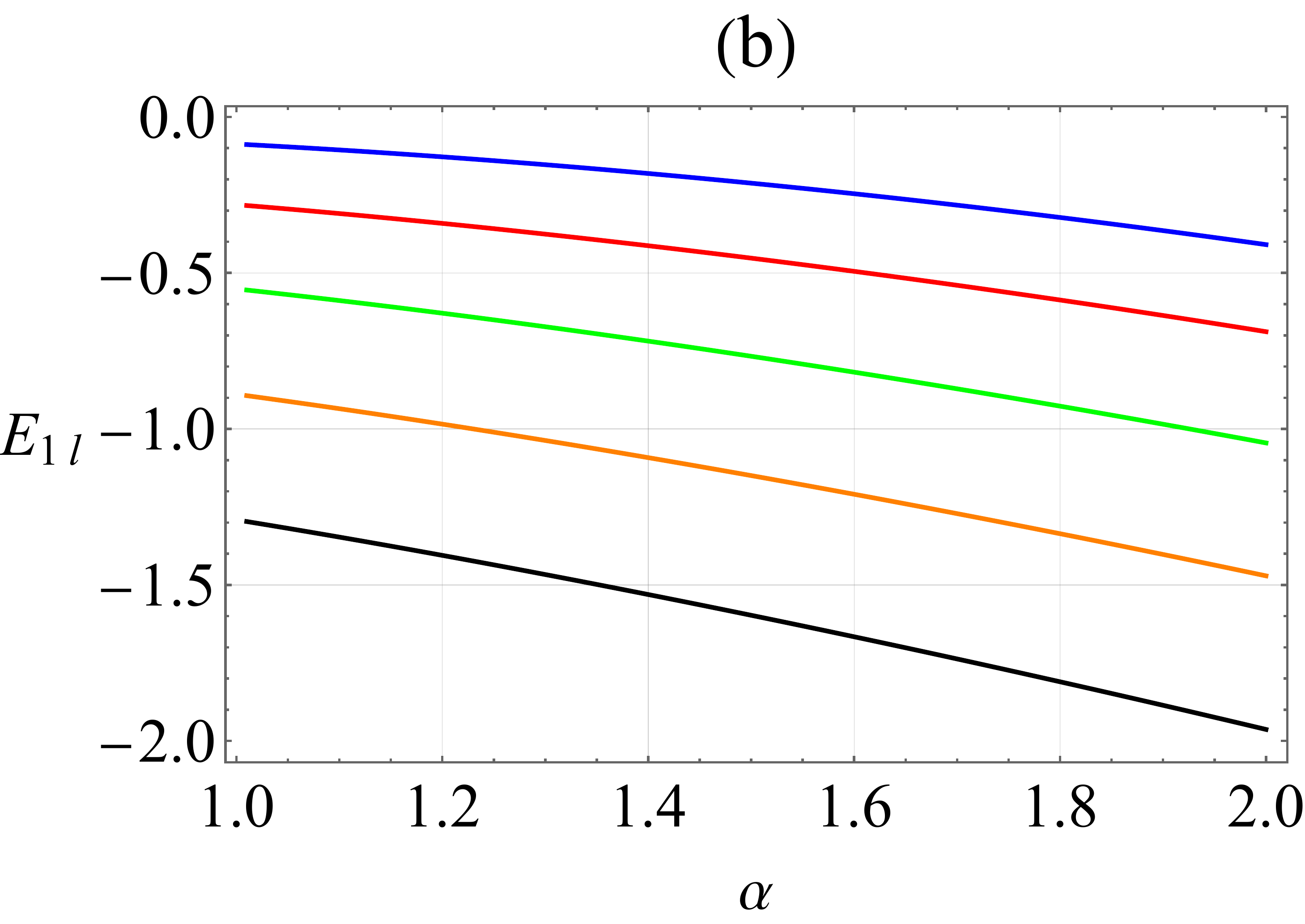}
	\caption{Energy levels (Equation (\ref{enexp})) with $n=1$ as a function of $\alpha$ for (a) $\xi=0.01$ and (b) $\xi=0.5$.}
	\label{Energy_alpha}
\end{figure}
Figure \ref{Energy_alpha} shows plots of energy levels as a function of $\alpha$ for $n=1$ and two different values of $\xi$. Since the parameter $\xi$ refers to the screening parameter of the Hulth\'{e}n potential, it must control the profile of these energy levels for some particular choice of the other parameters. As we can notice, $\xi=0.01$ (Fig. \ref{Energy_alpha}(a)) and $\xi=0.5$ (Fig. \ref{Energy_alpha}(b)) produce different plots, which correspond to distinct physical situations. In the case shown in Fig. \ref{Energy_alpha}(a), we see that for values of $\alpha$ in the range of 1.1 to 1.4, there is a decreasing tendency followed by an increase in $|E_{1l}|$. For $\alpha>1.4$, we see that $|E_{1l}|$ decreases. On the other hand, Fig. \ref{Energy_alpha}(b) reveals that $|E_{1l}|$ increases as $\alpha$ is increased.
\begin{figure}[!t]
	\centering
	\includegraphics[scale=0.32
	]{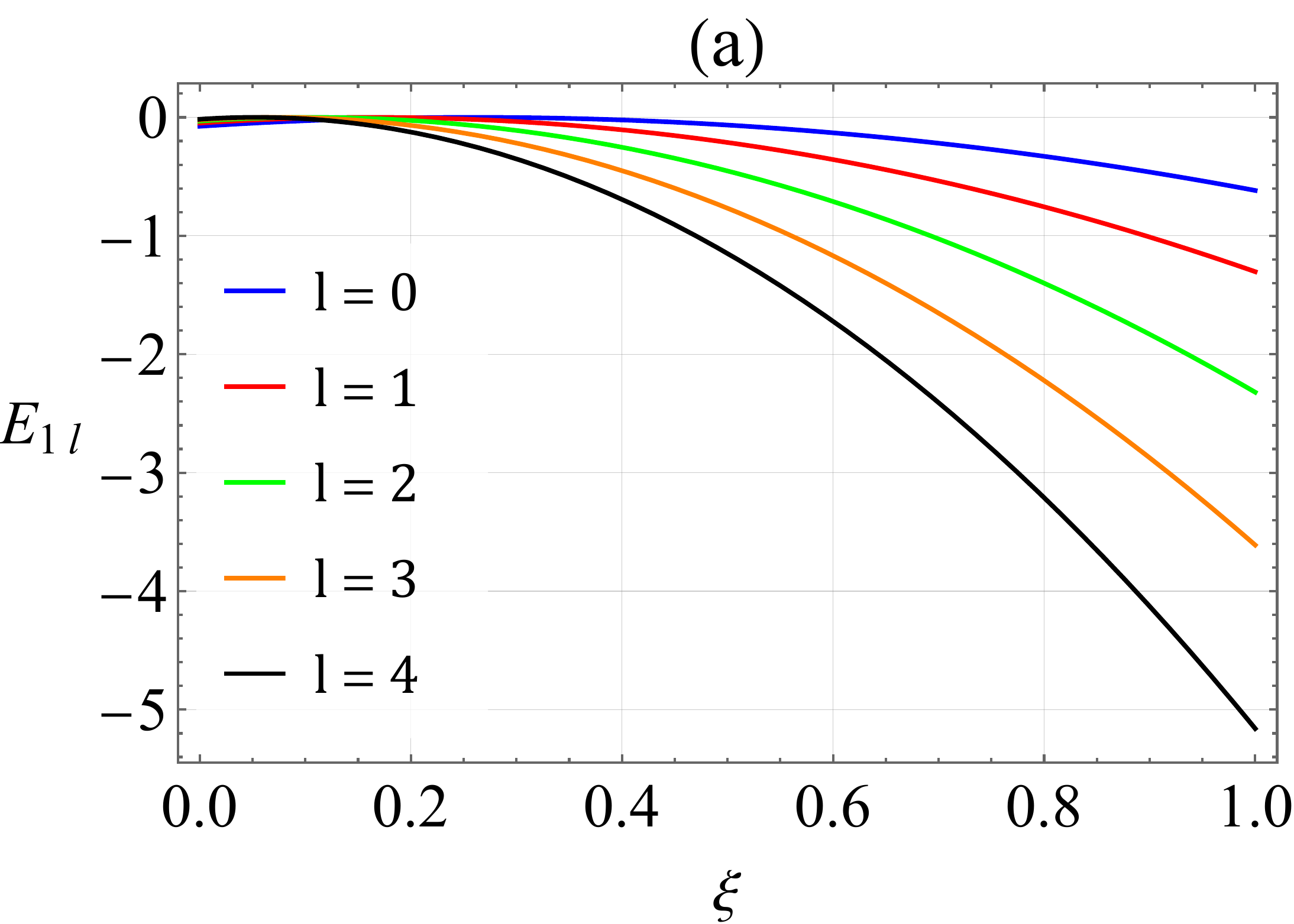}\qquad
	\includegraphics[scale=0.34
	]{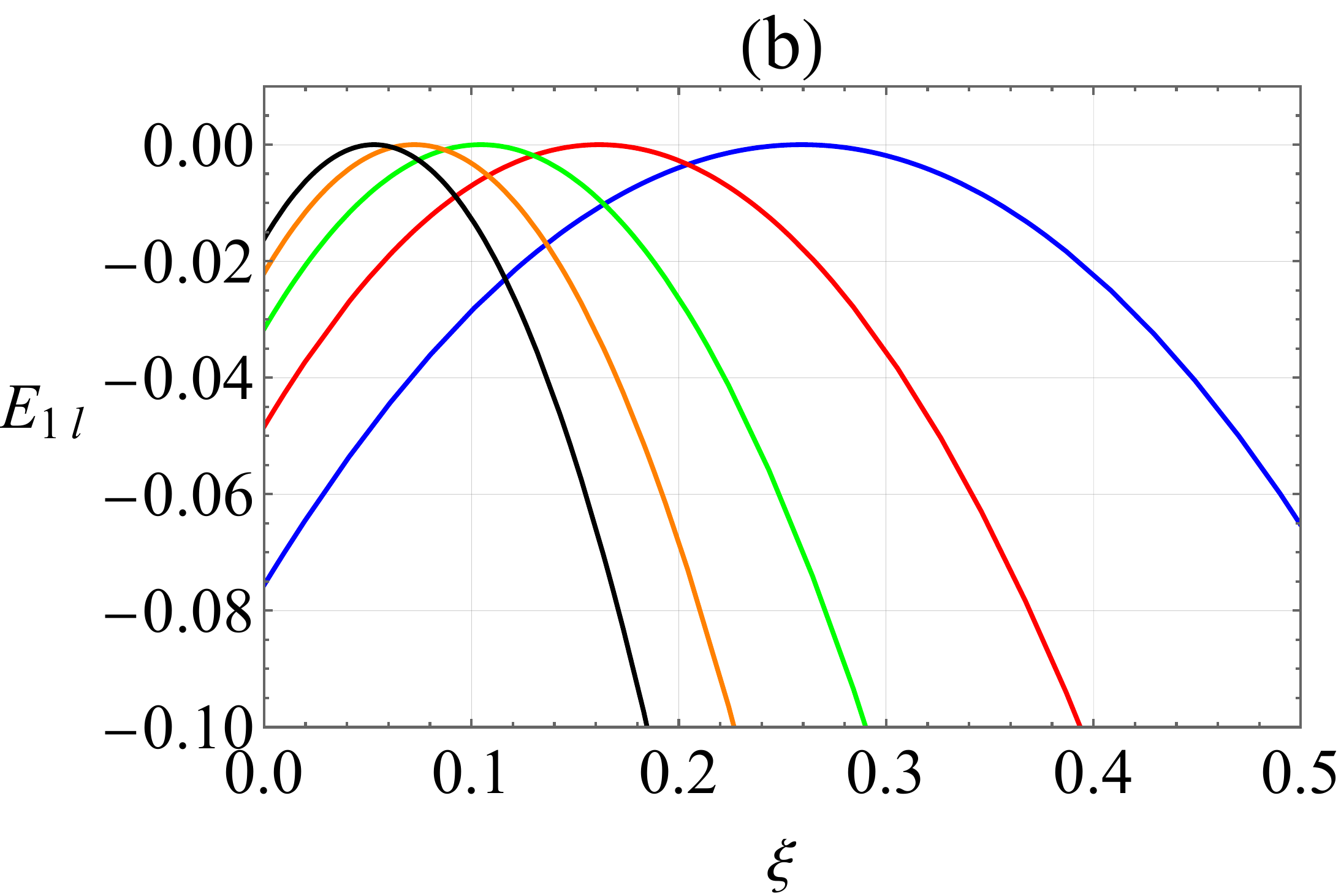}
	\caption{Energy levels (Equation (\ref{enexp})) with $n=1$ as a function of $\xi$ for $\alpha=1.5$. In (b), we show the range from (a) to $\xi=0.5$.}
	\label{Energy_xi}
\end{figure}

We also investigate how the energy levels are modified by considering different choices for the parameter $\xi$. In Figure \ref{Energy_xi}, we plot the energy levels $E_{1l}$ as a function of $\xi$ for the particular case when $\alpha=1.5$. As we can see, there is a range of $\xi$ where $|E_{1l}|$ decreases while for the other values $|E_{1l}|$ increases. For increasing values of $l$, $|E_{1l}|$ also increases (see e.g. the energy curve with $l=4$ (solid black curve) in Fig. \ref{Energy_xi}(a)). The various energy profiles illustrated in Figures \ref{Energy_l}, \ref{Energy_alpha} and \ref{Energy_xi} can be more easily interpreted when we make a reading of Figures \ref{Fig_veff} and \ref{Fig_veff2}. The energy intervals where $|E_{1\ell}|$ becomes small correspond precisely to the localized regions where a potential well begins to emerge. $|E_{1\ell}|$ can increase or decrease by adjusting the parameters involved.

The energy (\ref{enexp}) can be compared with other models in the literature. For example, if we take $\alpha=1$, we obtain the energies
\begin{equation}
	E_{nl}=-\frac{\hbar ^{2}\xi ^{2}}{2M}\left[ \frac{\frac{2MZe^{2}}{\hbar
			^{2}\xi }}{2\left( n+l+1\right) }-\frac{n+l+1}{2}\right] ^{2}.\label{bsep}
\end{equation}
Moreover, by making the changes $\xi=\delta$ and $M=\mu$ in Equation (\ref{bsep}), we find exactly the expression (24) of Reference \cite{PLA.2008.372.4779}, which also coincides with Equation (32) of Reference \cite{JPA.2006.39.11521}. 

\section{Conclusions}
\label{sec5}

In the present manuscript, we have investigated the problem of the quantum motion of an electron in the presence of the Hulth\'{e}n potential in the global monopole spacetime. Due to the existence of the global monopole background, it is necessary to consider the arising of a self-interaction potential. Under such conditions, we started from the Schrödinger equation with vector coupling in spherical coordinates. Then, we obtained the corresponding radial equation through the standard procedure in the literature. 
By analyzing several profiles of the effective potential, we have verified that the problem can be solved for both bound and scattering states. We confirm this by sketching profiles of the effective potential as a function of $r$ for different choices of the $\alpha$ parameter and considering some particular values for the parameter $\xi$. We have used the exponential function transformation approach and an approximation for the centrifugal potential to transform the radial equation coming from the Schr\"{o}dinger equation into a differential equation of the hypergeometric type. We have solved this equation for scattering states and found an expression for the phase shift. We adopted the following procedure to find an expression for the bound state energies: we first obtained the $S$-matrix. Then, we analyzed its poles. We examined the profile of energies considering the situation where the self-interaction potential is attractive and repulsive. Through some sketches, we showed that both this potential and the $\alpha$ parameter could modify the bound state energies. We have also investigated the probability density function and its dependence on the parameters $\alpha$, $\xi$, and the quantum number $n$. Alternatively, we also solve the problem for bound states using the Frobenius method and confirm that the bound state energies and wave functions are the same as those already obtained. The justification for using the Frobenius method is that it provides us with an expression that establishes a condition for the occurrence of bound states (Equation (\ref{cvl})).

\section*{Acknowledgments}

This work was partially supported
by the Brazilian agencies CAPES, CNPq, and FAPEMA. E. O. Silva acknowledges CNPq
Grant 306308/2022-3, FAPEMA Grants PRONEM-01852/14 and UNIVERSAL-06395/22. M. M. Cunha acknowledges CAPES Grant 88887.358036/2019-00.
This study was financed in part by the Coordena\c{c}\~{a}o de
Aperfei\c{c}oamento de Pessoal de N\'{\i}vel Superior - Brasil (CAPES) -
Finance Code 001.

\bibliographystyle{apsrev4-2}

\end{document}